\documentclass[3p,final,12pt]{elsarticle}
\usepackage{graphicx}
\usepackage{epstopdf}
\usepackage{float}
\usepackage{booktabs}
\usepackage{caption}
\usepackage{subcaption}
\usepackage{natbib}
\usepackage{tabularx}
\usepackage{lipsum}  
\usepackage{amsmath}
\usepackage{amsthm}
\newtheorem{lemma}{Lemma}
\usepackage{amssymb}
\usepackage[symbol]{footmisc}
\usepackage{hyperref}

\journal{Physics of Fluids}

\begin{document}

\begin{frontmatter}



\title{Applied Neural Network-Based Active Control for Vortex-Induced Vibrations Suppression in a Two-Degree-of-Freedom Cylinder\footnote[2]{This is the accepted manuscript reprinted with permission from Soha Ilbeigi, Ashkan Bagherzadeh, Alireza Sharifi; Applied neural network-based active control for vortex-induced vibrations suppression in a two-degree-of-freedom cylinder. Physics of Fluids 1 January 2025; 37 (1): 017150. \hyperlink{https://doi.org/10.1063/5.0251228}{https://doi.org/10.1063/5.0251228.} Copyright 2025, AIP Publishing. This article may be downloaded for personal use only. Any other use requires prior permission of the authors and the AIP Publishing.}}


\author[inst1]{Soha Ilbeigi}
\author[inst1]{Ashkan Bagherzadeh\footnote[3]{This author is currently pursuing a Ph.D. at Michigan State University. This
research was conducted during his MSc studies at Sharif University of Technology.}}
\author[inst1]{Alireza Sharifi}

\affiliation[inst1]{organization={Department of Aerospace Engineering, Sharif University of Technology},
            addressline={Azadi Street}, 
            city={Tehran},
            postcode={1458889694}, 
            state={Tehran},
            country={Iran}}



\begin{abstract}
Vortex-Induced Vibrations (VIVs) of cylindrical structures present significant challenges in various engineering applications, including marine risers, tall buildings, and renewable energy systems. Hence, it is vital to control Vortex-Induced Vibrations of cylindrical structures. For this purpose, in this study a novel approach is introduced to VIV control, based on a model-based active control strategy integrated with a Neural Network (NN) in the presence of uncertainty modeling.
The proposed method utilizes a closed-loop control system, where feedback from the system's dynamic state is used to generate adaptive control commands, enabling the system to respond to changing flow conditions and nonlinearities. Then, the controllability analysis is conducted to assess the efficiency of the control strategy in mitigating VIV. Two control approaches are implemented: simple learning and composite learning. Both strategies significantly enhance vibration suppression, achieving up to 99\% reduction in vibrations despite uncertainties in the system. The results demonstrate the potential of the proposed method to enhance the efficiency, stability, and lifespan of structures subject to VIV.
\end{abstract}


\end{frontmatter}


\section{Introduction}
\label{sec: introduction}

The VIV phenomenon occurs when fluid flowing past a quasi-cylindrical structure generates oscillations in the structure due to the shedding of vortices on either side, creating forces perpendicular to the fluid flow \cite{Peng2023}. Circular cylinders are among the most widely studied geometric forms in this area, given their common applications in engineering \cite{Zdravkovich1997}, such as in marine risers \cite{Krishnakumari2023}, bridge pylons \cite{sanchez2015vortex}, and floating platforms \cite{Yin2022}. The study of VIV in cylindrical structures is crucial because these oscillations can cause structural damage or even failure \cite{Lu2022}, especially in structures subjected to fluctuating fluid flows. In the renewable energy sector, particularly in ocean currents and tidal turbines, effective VIV damping enhances system efficiency and durability \cite{Yin2022}, and in applications like the design of tall buildings and chimneys \cite{Christensen2017}, where aerodynamic and hydrodynamic stability is critical, understanding VIV is essential. Damping these vibrations can significantly extend the lifespan of structures, reduce maintenance costs, and improve safety. To tackle these challenges, diverse control strategies, called VIV control, have been designed to reduce VIV and improve the structural efficiency of cylindrical systems.

VIV control strategies can be broadly categorized into passive and active control. Passive control typically involves modifying the flow through control surfaces or geometric adjustments \cite{Kumar2008}, irrespective of current conditions. While passive methods are cost-effective and simpler, they lack adaptability to changing flow conditions, leading to a shift toward active control methods \cite{king2010active}. In contrast, active control uses real-time sensor feedback to apply external forces that adjust flow and vortices, proving more effective under unpredictable conditions.

Active control can be further categorized into two main strategies: flow control and structural vibration control. Flow control methods manipulate the fluid environment, for instance, steady suction/blowing \cite{Chen2013}, synthetic jets \cite{Wang2016}, electromagnetic forcing \cite{Zhang2014}, rotation \cite{Zhu2017}, thermal effects and acoustic excitation \cite{Wan2016}, and traveling wave wall as a flexible surface \cite{Xu2014-travelingwall}, whereas structural vibration control directly targets the structural response using actuators or feedback mechanisms.

When addressing structural vibration control, a distinction can be made between open-loop and closed-loop systems. Open-loop systems operate without feedback from the structure's response, relying solely on pre-defined control signals. In contrast, closed-loop systems continuously monitor the structural vibrations and dynamically adjust control inputs based on real-time feedback. Open-loop control can effectively suppress vortex shedding and structural vibrations, provided the actuating signal is precisely tuned to match the system's natural frequency. However, achieving optimal performance with open-loop control requires relatively large actuator perturbation amplitudes, and the perturbation frequency must remain within a narrow range around the resonance frequency \cite{Hasheminejad2014}. These constraints limit the adaptability of open-loop systems to varying operating conditions, a challenge that closed-loop systems are specifically designed to overcome through their responsive and adaptive capabilities.

Among closed-loop strategies, model-free and model-based approaches are widely employed. Model-free control, or blind control, includes methods like Proportional-Integral-Derivative (PID) controllers. For instance, \citet{Zhang2004-PID} demonstrated that a PID controller could effectively mitigate vortex-induced vibrations in a flexibly supported square cylinder by optimizing the control of vortex shedding dynamics. \citet{mohammed2014fuzzyPID} further advanced this approach by developing a Fuzzy-PID controller, which demonstrated superior performance compared to traditional PID methods by utilizing Adaptive Neuro-Fuzzy Inference Systems (ANFIS) and neural network-based models for enhanced system identification and response. Similarly, \citet{Hasheminejad2023} proposed an intelligent self-tuning fuzzy PID algorithm, which adaptively adjusts parameters based on real-time system feedback to achieve effective vibration suppression in complex flow environments. 

Despite their simplicity, model-free approaches often fail to handle the nonlinear dynamics, disturbances, and modeling errors inherent in VIV systems. Deep Reinforcement Learning (DRL) methods have been proposed to address these limitations by involving experimental models. While they are not purely model-based (as the model-based category is defined by the use of a mathematical model of the system), these approaches can be considered model-free. For example, \citet{Ren2019} applied GP to adjust blowing/suction jets, achieving 94.2\% VIV suppression with superior energy efficiency. \citet{Zheng2021} deployed 152 velocity sensors around a vibrating cylinder for feedback, achieving 82.7\% VIV suppression. However, setup complexity and sensor requirements remain limitations. \citet{Ren2024} demonstrated that DRL methods exhibit considerable potential in optimizing performance and improving adaptability to varying conditions in VIV control. However, while these model-free strategies are flexible, their inability to leverage detailed system modeling often limits their precision and robustness, making model-based control a more reliable alternative for achieving optimal vibration suppression.

In model-based approaches, \citet{Martin1976} used pole placement algorithms to address structural vibrations, emphasizing the long-standing use of linear control for structural response mitigation. \citet{Yang1995} introduced Sliding Mode Control (SMC), a robust control technique that offers excellent stability and can handle uncertainties. However, SMC can lead to issues like chattering, which may activate unmodeled high-frequency dynamics. To address this, SMC can be combined with other techniques like fuzzy control. \citet{Symans1999} noted that Fuzzy Logic Control (FLC) could be a useful tool for structural vibration suppression, especially for handling nonlinear systems that traditional methods may not effectively control. However, conventional FLC lacks adaptability since it typically requires predefined parameters. \citet{Hasheminejad2013} applied Linear Quadratic Gaussian (LQG) to control VIV, demonstrating the effectiveness of model-based techniques in vibration suppression. \citet{Hasheminejad2014} implemented an Adaptive Fuzzy Sliding Mode Controller (AFSMC) to actively control VIVs of a circular cylinder at low Reynolds numbers, achieving substantial vibration suppression by addressing nonlinear flow-structure interactions. \citet{Lou2021} extended this concept to the active control of risers in the "lock-in" region, utilizing AFSMC to enhance the stability and performance of VIV control in harsh marine environments. Additionally, Kalman filter-based controllers, as studied by \citet{Yao2017}, have been shown to effectively stabilize unstable flows and suppress VIV by accurately estimating system states and integrating feedback into the control loop. 

However, these model-based methods typically do not explicitly consider uncertainties or disturbances in their design, or if considered, they often assume these uncertainties to be fully known or precisely characterized for the controller. This limitation reduces their effectiveness in real-world scenarios where uncertainties due to flow nonlinearity, structural variability, or actuator faults are inherently unknown. This underscores the need for novel strategies, such as neural network-based controllers, which hold promise for addressing these challenges comprehensively. Neural Networks, with their powerful function approximation capabilities, provide a promising solution by adaptively estimating unknown terms. By integrating neural networks into the control framework, it is possible to dynamically estimate and compensate for these uncertainties, improving the robustness and performance of the control system.

This research proposes a novel approach, called NN-VIV Suppression, based on a neural network to estimate the unknown uncertainty term within the control model for a two-degree-of-freedom cylinder. Building on the model-based category, this study focuses on an active, model-based control strategy that integrates real-time feedback and neural network-based uncertainty estimation to optimize vibration suppression. The proposed control approach relies on a closed-loop system, where dynamic feedback from the system's state is continuously used to adjust control commands, allowing for real-time adaptation to further changing flow conditions. A comprehensive controllability analysis is performed to evaluate the ability of the control system to effectively influence the structure's vibrations. This analysis ensures that control inputs can be manipulated to suppress VIV efficiently under various operating conditions. The neural network estimates the unknown uncertainty, which is inherently unobservable by the controller. This approach not only demonstrates the potential to improve vibration control but also ensures the stability and performance of structures exposed to vortex-induced vibrations. 

The following sections will discuss the problem definition in Sec.~\ref{sec: Problem Definition}, modeling in Sec.~\ref{sec: Modeling of the Cylinder}, uncertainty estimation in Sec.~\ref{subsec: NN arch}, controller design in Sec.~\ref{sec: controller formulation}, results in Sec.~\ref{sec: results}, and conclusions in Sec.~\ref{sec: conclusion}.

\section{Problem Definition}
\label{sec: Problem Definition}

\begin{figure}[H]
  \centering
  \includegraphics[width=1\linewidth]{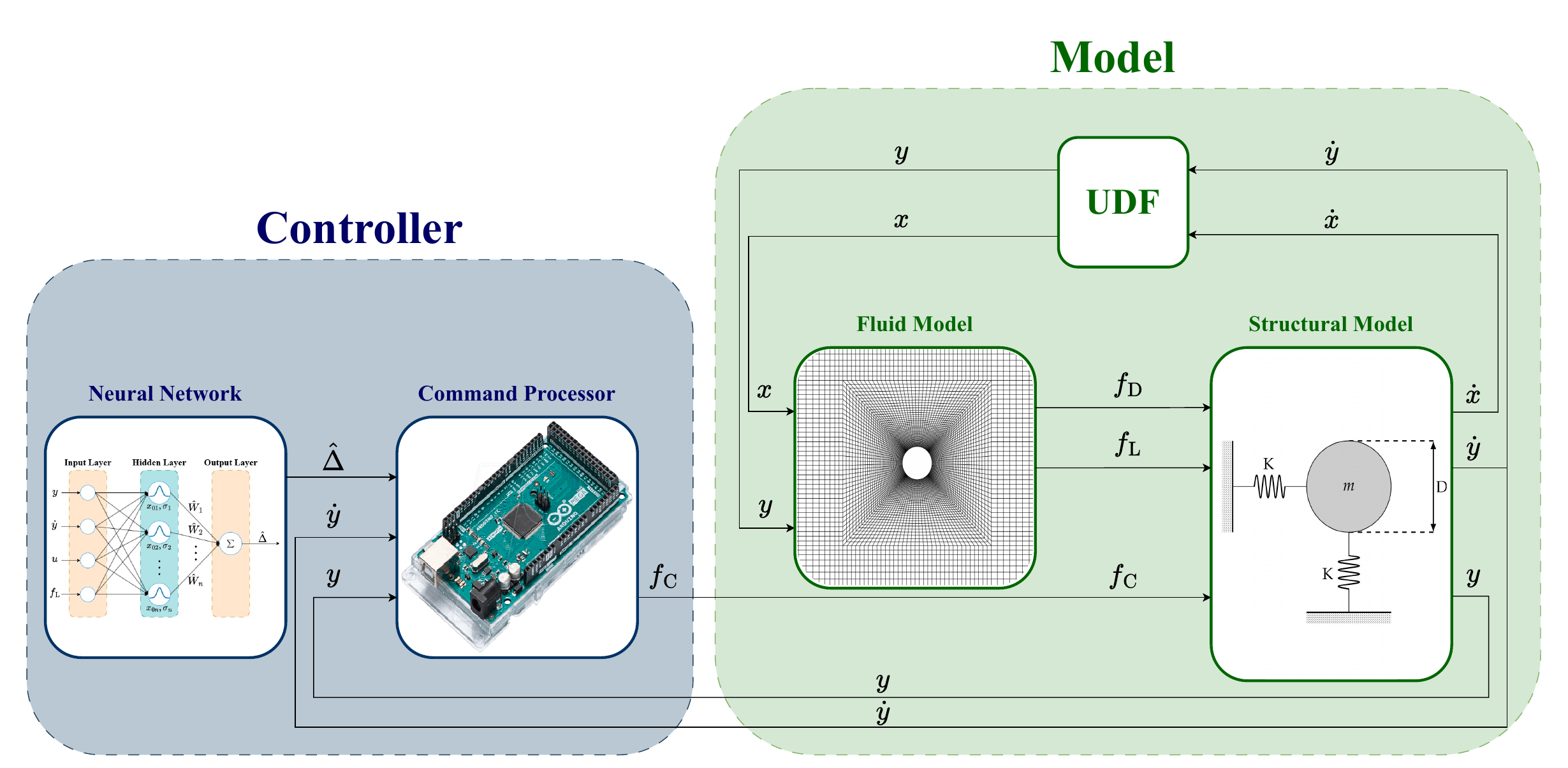}
  \caption{Structure of the NN-VIV Suppression simulation method}
  \label{coupling}
\end{figure}

The simulation procedure for NN-VIV suppression of the cylinder is schematically illustrated in Fig.~\ref{coupling}. This architecture includes:
(1) the fluid model of the cylinder, implemented in ANSYS/Fluent software, to compute the aerodynamic forces of lift and drag,
(2) the structural model of the cylinder, implemented in MATLAB/Simulink environment, to compute the cylinder's acceleration, velocity, and position in the flow direction (\(x\)) and perpendicular to the flow (\(y\)) for the next time step,
(3) a User-Defined Function (UDF), transferring the necessary information between the fluid and structural models at each time step in order to complete the Fluid-Solid Interaction (FSI) loop,
(4) an accelerometer sensor for measuring the deflection of the cylinder,
(5) the estimated uncertainty using the neural network to utilize in the control command,
(6) controller for providing the control force to suppress the vibrations of the structural model.
The steps of this block diagram are denoted in the following sections.

\section{Modeling of the Cylinder Dynamics}
\label{sec: Modeling of the Cylinder}

\begin{figure}[H]
  \centering
  \includegraphics[width=0.8\linewidth]{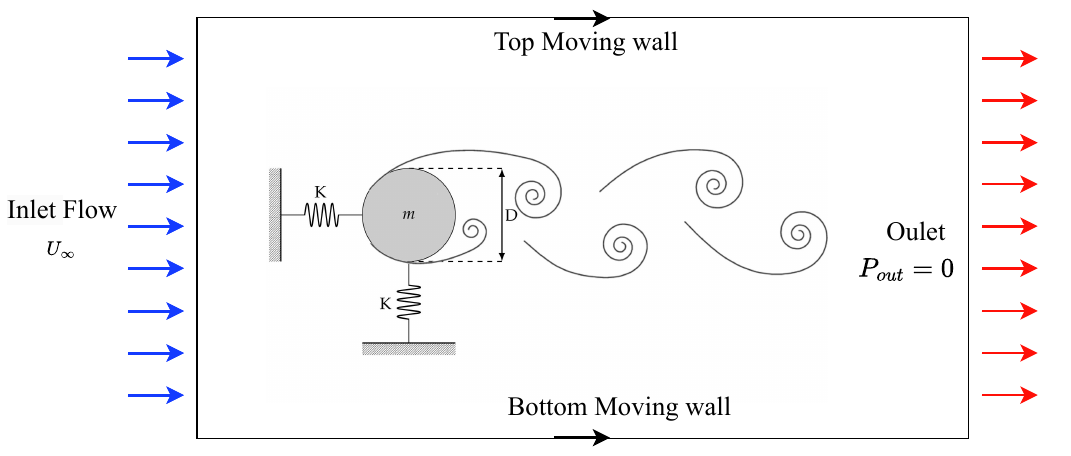}
  \caption{Structural model within the fluid solution domain}
  \label{sol domain}
\end{figure}

In order to design an effective control system for a two-degree-of-freedom (2DOF) cylinder, it is essential to establish an accurate dynamic model. The dynamic behavior of the system is influenced by both the fluid forces acting on the cylinder and the cylinder's structural dynamics. Fig.\ref{sol domain} illustrates the schematic view of the fluid domain and the placement of the structural model within this domain. In the following, two distinct models, fluid and structural, are formulated and discussed in detail.

\subsection{Fluid Model}
\label{subsec: Fluid Model}

\begin{figure}[H]
     \centering
     \begin{subfigure}[b]{0.36\textwidth}
         \centering
         \includegraphics[width=\textwidth]{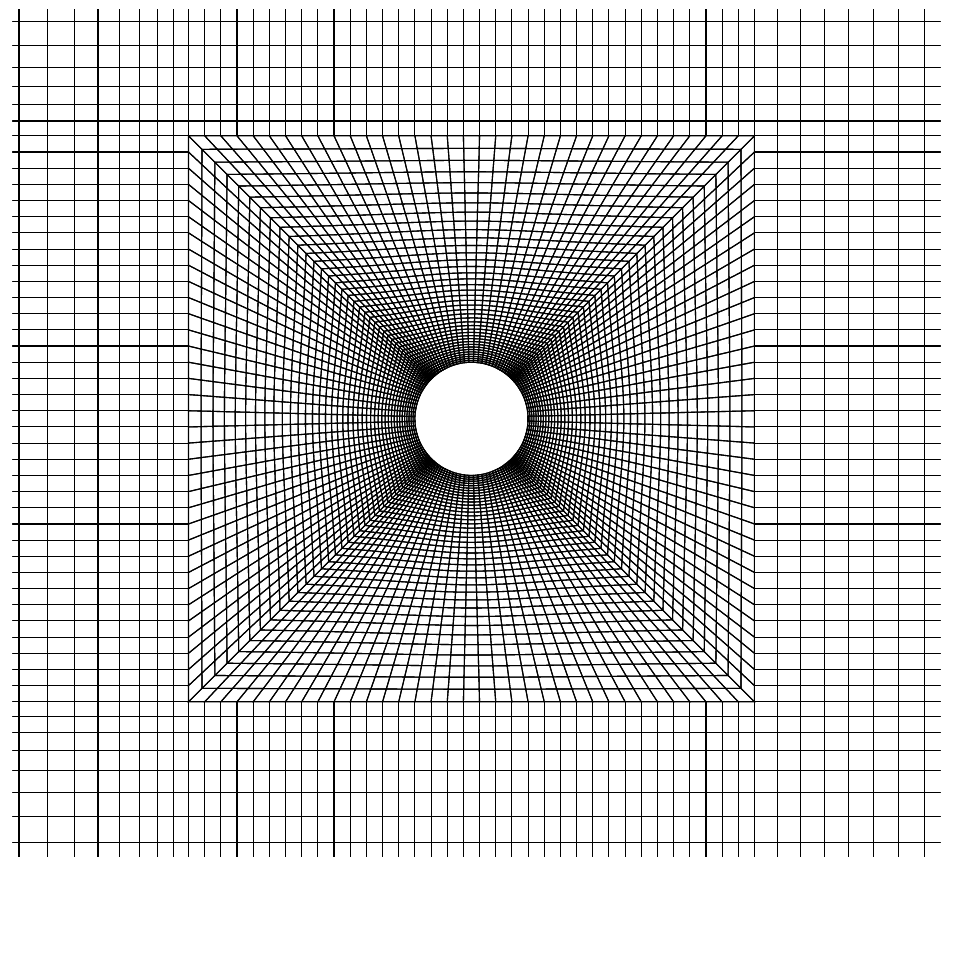}
         \caption{Grid density distribution near the cylinder}
         \label{figa_mesh}
     \end{subfigure}
     \hfill
     \begin{subfigure}[b]{0.6\textwidth}
         \centering
         \includegraphics[width=\textwidth]{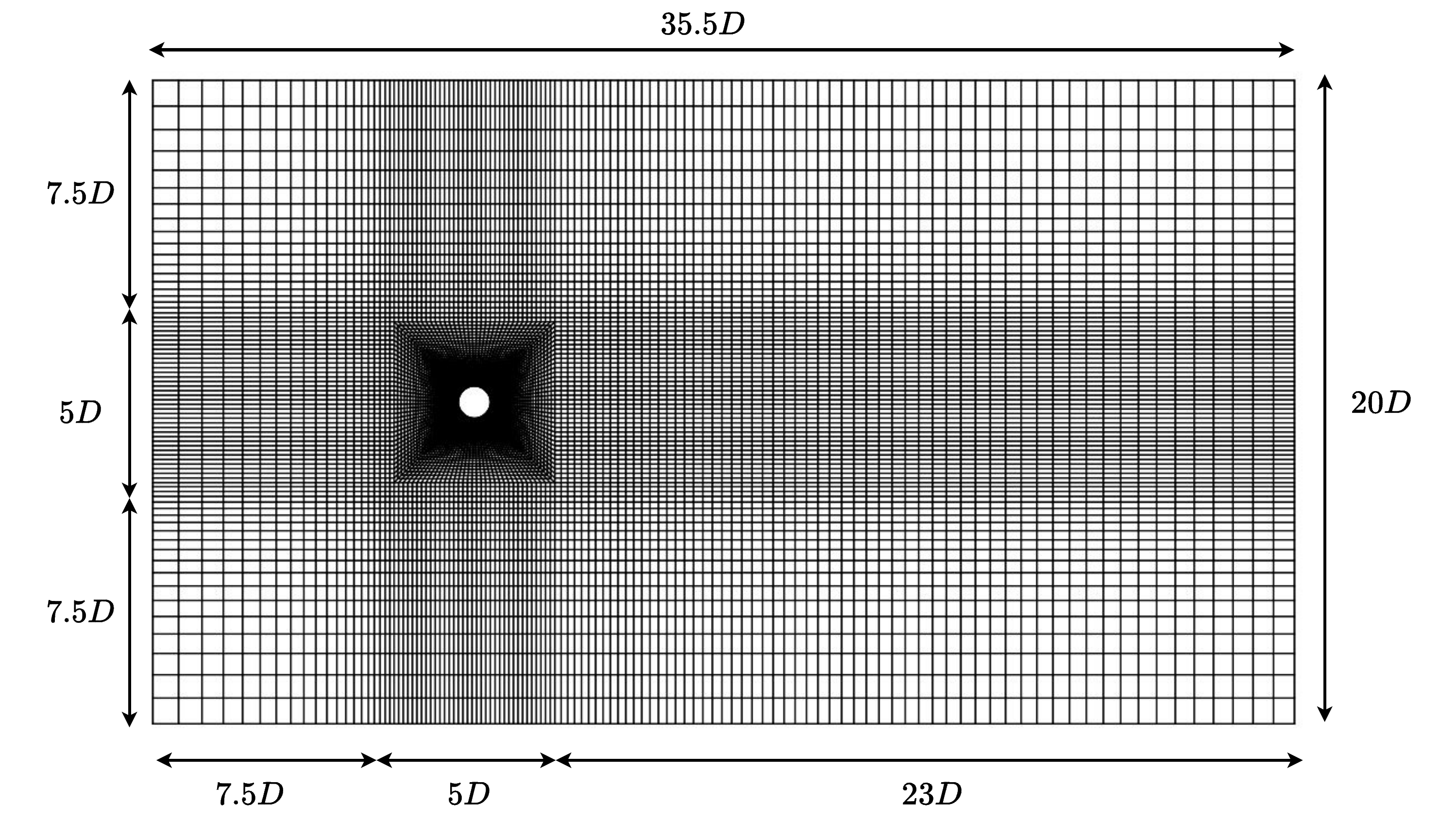}
         \caption{Dimensions of the computational domain}
         \label{figb_mesh}
     \end{subfigure}
        \caption{Computational domain of fluid model}
        \label{figmesh}
\end{figure}

In this section, first, the specifications of the computational grid are introduced. Then, the governing equations and numerical approaches to solve them are presented. Fig.\ref{figmesh} depicts the computational grid around the cylinder. As illustrated, the computational grid consists of two zones: (a) a square region around the cylinder with mapped structured grids, and (b) structured meshes around the square zone in the rectangular domain. 

This generated mesh, characterized by its ability to undergo deformation, can alter its shape in response to cylinder movement and boundary conditions. In the employed moving mesh strategy, at each time step, the moving/deforming function updates the meshes for calculations within the time step. Due to the requirement for higher accuracy in the solution within the square region surrounding the cylinder, the number and density of mesh cells are significantly greater compared to those in the far field. The total number of structured cells selected after the completion of the mesh independence test is 43,000, with no alteration in the results observed for finer meshes. The same independence study procedure was applied to determine the optimal time step size necessary for the simulation, which equated to $\Delta t = 0.2$ s.

Considering the unsteady, viscous, and incompressible nature of the fluid flow, characterized by constant fluid properties, the modeling approach utilizes the Unsteady Reynolds-Averaged Navier–Stokes (URANS) equations. These equations encompass both mass and momentum conservation equations and are employed to simulate the flow around the cylinder.

\begin{equation}
     \nabla \cdot \mathbf{u} = 0,
\end{equation}

\begin{equation}
    \frac{\partial \mathbf{u}}{\partial t} +    (\mathbf{u}\cdot \nabla)\cdot \mathbf{u}= -\nabla p + \frac{1}{Re}\nabla^2 \mathbf{u}.
\end{equation}

In this context, $\mathbf{u}$ represents the flow velocity vector, $t$ signifies time, $p$ denotes pressure, and $Re$ is the Reynolds number of the flow around the cylinder, expressed as $Re=\ U\text{D}/\nu$, where $U$ represents the free stream flow velocity, $\text{D}$ signifies the characteristic length (cylinder diameter), and $\nu$ stands for kinematic fluid viscosity. 

The Finite Volume Method (FVM) was used to discretize the aforementioned governing equations, utilizing the following spatial and temporal discretization methods for each variable to achieve accurate and stable solutions. Pressure employed a standard first-order accurate spatial scheme, while momentum utilized the QUICK scheme (Quadratic Upwind Interpolation for Convective Kinematics) with second-order accuracy. Time derivative terms were discretized using a first-order implicit formulation. Additionally, the SIMPLE (Semi-Implicit Method for Pressure Linked Equations) algorithm was implemented for pressure-velocity coupling. 
Finally, to validate the presented numerical model, the results are discussed comprehensively in Sec.~\ref{subsec: eval fluid model}.

\subsection{Structural Model}
\label{subsec: Structural Model}

For modeling the structural part of the system, the cylinder is treated as a 2DOF structure, with one degree of freedom along the flow direction and the other perpendicular to the flow, as illustrated in Fig.\ref{sol domain}. To accurately represent this model and the interaction between the structural and fluid components, the equations of motion of the cylinder are derived based on Newton’s second law as follows:

\begin{equation} \label{structural eq}
    \left[ \begin{array}{c}
    \ddot{x} \\
    \ddot{y}
    \end{array} \right] = \frac{1}{m} \left[ \begin{array}{c}
    f_{\rm{D}} - \text{K} x \\
    f_{\rm{L}} - \text{K} y
    \end{array} \right] + \frac{1}{m} \left[ \begin{array}{c}
    0 \\
    f_{\rm{C}}
    \end{array} \right],
\end{equation}

where $m$ and $\text{K}$ represent the cylinder's mass and spring stiffness, respectively. Additionally, $f_{\rm{D}}$ and $f_{\rm{L}}$ are the fluid forces acting on the cylinder, where $f_{\rm{D}}$ denotes the drag force in the flow direction, while $f_{\rm{L}}$  indicates the lift force acting perpendicular to the flow. Finally, $f_{\rm{C}}$ is the control command, applied perpendicular to the flow to dampen the oscillations.
\section{Uncertainty Estimation}
\label{subsec: NN arch}

This section presents the architecture of the Neural Network implemented for uncertainty estimation. The design and functionality of the Neural Network are detailed below.

\subsection{Neural Network Structure}
\label{subsec: NN_structure}

\begin{figure}[H]
  \centering
  \includegraphics[width=0.7\linewidth]{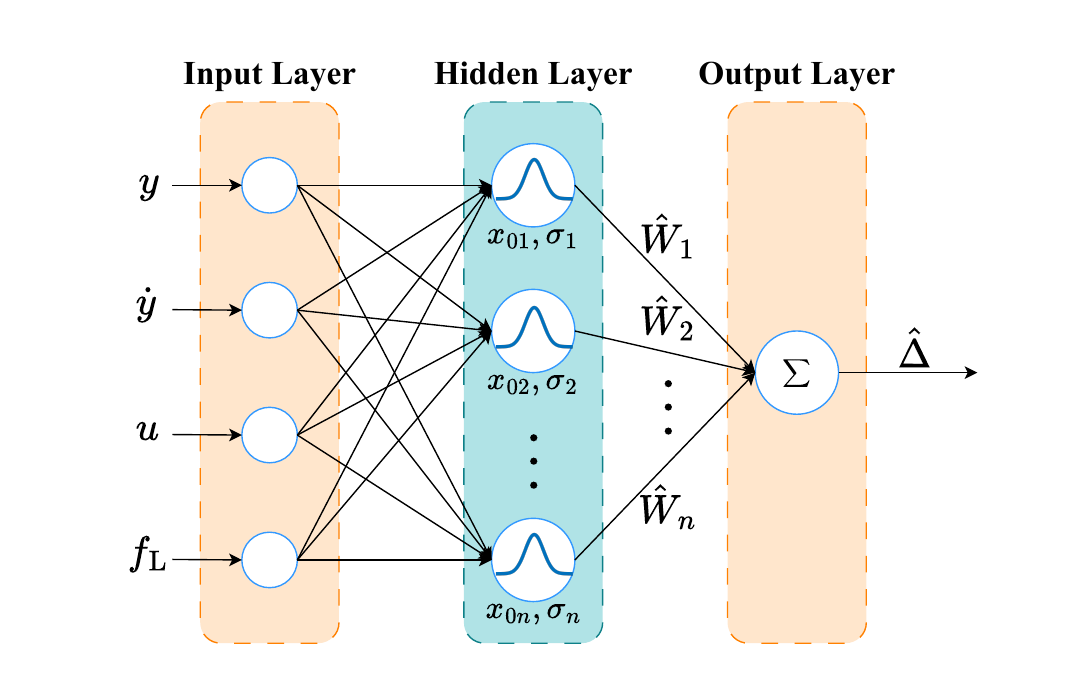}
  \caption{Structure of the proposed neural network}
  \label{NN}
\end{figure}

As shown in Fig.~\ref{NN}, the Neural Network consists of three layers: Input, Hidden, and Output layers. In the current study, the Neural Network inputs are defined as: 

\begin{equation} \label{Xin}
X = \left[ {\begin{array}{*{20}{c}}y&{\dot y}&u&{{f_{\text{L}}}}\end{array}} \right]^{T}
\end{equation}

\subsection{Hidden Layer and Activation Function}
\label{subsec: hidden_layer}

The Hidden Layer contains $n$ neurons, with the Radial Basis Function (RBF) as their activation function \cite{Liu2013-rbf}. The formulation of the RBF is given by:
\begin{equation} \label{RBF eq}
     \mu \left( y \right) = \exp \left( { - \frac{{\left\| {X - {x_{0i}}} \right\|^2}}{2\sigma _i^2}} \right), \quad i = 1, 2, 3, \ldots ,n,
\end{equation}
where $\sigma _i$ is a positive scalar (the width), and $n$ denotes the number of hidden nodes (neurons). Each hidden node has a center vector ${x_{0i}}$, which is a parameter vector of the same dimension as the input vector $X$. The parameters $\sigma _i$ and ${x_{0i}}$ for each node are initialized randomly.

\subsection{Output Layer and Approximation}
\label{subsec: output_layer}

The Neural Network leverages the universal approximation property to estimate the unknown vector \( \Delta \). The relationship is defined as:
\begin{equation} \label{Delta eq}
     \Delta = W^T \mu(y) + \varepsilon,
\end{equation}
where $W$ denotes the unknown optimal weights, and $\varepsilon$ is the bounded approximation error ($\left\| \varepsilon \right\| \le \varepsilon_M$). 

Since \( \Delta \) is unknown, it is approximated using the Neural Network as:
\begin{equation} \label{Deltahat eq}
     \hat \Delta  = {\hat W^T}\mu \left( y \right),
\end{equation}
where ${\hat W}$ represents the output layer weights, and $\hat \Delta$ denotes the network output. Here, $\mu \left( y \right)$ is a $n \times 1$ matrix, and ${\hat W}$ is also $n \times 1$, resulting in $\hat \Delta$ being a scalar.

\subsection{Universal Approximation and Convergence}
\label{subsec: universal_approximation}

To approximate an unknown target function \( \Delta \), the universal approximation theorem is utilized, which states that a sufficiently large neural network can approximate any continuous function on a compact set to arbitrary precision \cite{Hornik1989}. 

Assuming the true model for \( \Delta \) is given by:
\begin{equation}
    \Delta = W^T \mu(y),
\end{equation}
the network is trained to minimize the approximation error. Under the universal approximation property, the estimated weight matrix \( \hat{W} \) converges to values closely approximating \( W \). The Neural Network's output is discussed and evaluated in Sec.~\ref{subsec: eval nn}.

\section{Formulation of Controller Design} \label{sec: controller formulation}

This section presents the design and implementation of a Neural Network-based controller. The Feedback Error Learning (FEL) method, widely recognized for its integration of Neural Networks into adaptive control design \cite{Emami2022-na}, is employed in this study. FEL facilitates the seamless combination of control design with online parameter updates of the Neural Network, enabling effective compensation for model uncertainties and external disturbances. Typically, the control system consists of an inner-loop conventional controller that stabilizes the system dynamics, while the Neural Network-based controller provides auxiliary compensation for uncertainties. By using a composite Lyapunov function that includes both the tracking error and the Neural Network parameter estimation error, the closed-loop system satisfies the Bounded-Input–Bounded-Output (BIBO) stability criterion in the presence of uncertainties and disturbances.
\subsection{Dynamical System}
\label{subsec: dynamical system}

The vertical axis of the structural model exclusively considered for control, as defined in Sec.~\ref{subsec: Structural Model}, rewritten in affine form as follows:
\begin{equation} \label{affine form}
     \ddot y = f\left( y, \dot{y} \right) + bu + \Delta\left( y, \dot{y}, u \right),
\end{equation}
where $f\left( y, \dot{y} \right) = \frac{1}{m}\left( {{f_\text{L}} - \text{K} y} \right)$, $b = \frac{1}{m}$, $u = f_\text{C}$, and $\Delta(y, \dot{y}, u) = \Delta_1(y, \dot{y}) + \Delta_2(u) u + \Delta_3$ represents model uncertainties and external disturbances. For the sake of simplicity, in the following, $f\left( y, \dot{y} \right)$ and $\Delta(y, \dot{y}, u)$ will be denoted as $f$ and $\Delta$, respectively.
\subsection{Nonlinear Controllability Analysis}
\label{subsec: Controllability}

In the context of nonlinear systems, controllability analysis plays a crucial role in determining whether a system can be driven from one state to another using available inputs. Nonlinear controllability fundamentally differs from its linear counterpart, requiring more sophisticated mathematical tools due to the inherent complexities of nonlinear dynamics. Nonlinear control systems lack a general controllability criterion similar to that available for linear systems, necessitating specialized approaches, such as Lie algebraic methods, for comprehensive analysis. The following lemma \cite{Hermann1977} provides a foundation for analyzing the controllability of a nonlinear system using Lie bracket conditions, which assess controllability by examining the vector fields associated with the system's dynamics.

\begin{lemma}
\textbf{(Lie Bracket Criterion for Controllability)}

Consider a nonlinear dynamic system represented by
\[
\dot{x} = \mathbf{f(x)} + \mathbf{g(x)} u,
\]
where \( x \in \mathbb{R}^n \) is the state vector, \( u \in \mathbb{R}^m \) is the control input vector, and \( \mathbf{f(x)} \) and \( \mathbf{g(x)} \) are smooth vector fields on \( \mathbb{R}^n \). Define the controllability matrix \( M \) as
\[
M = \left[ \mathbf{g(x)}, [\mathbf{f(x)}, \mathbf{g(x)}], [\mathbf{g_i(x)}, \mathbf{g_j(x)}], [\mathbf{f(x)}, [\mathbf{f(x)}, \mathbf{g(x)}]], \ldots \right],
\]
where \( [\cdot, \cdot] \) denotes the Lie bracket of two vector fields. The system is locally controllable around a point \( x_0 \) if \( M \) has full rank at \( x_0 \), i.e., if
\[
\text{rank}(M) = n.
\]
\end{lemma}

Now, let us consider the nonlinear model defined by Eq.~\ref{affine form} and transform it into state-space form as follows:
\begin{equation} \label{state space}
    \begin{aligned}
        \dot{\mathbf{x}} &= \left[ \begin{array}{c} \dot{x_1} \\ \dot{x_2} \end{array} \right] = \left[ \begin{array}{c} x_2 \\ \frac{1}{m} \left( f_{\text{L}} - K x_1 \right) + bu + \Delta(\mathbf{x},u) \end{array} \right],
    \end{aligned}
\end{equation}
where \( 
\mathbf{x} = \begin{bmatrix} x_1 \\ x_2 \end{bmatrix} = \begin{bmatrix} y \\ \dot{y} \end{bmatrix}, \quad
\mathbf{f(x)} = \begin{bmatrix} x_2 \\ \frac{1}{m} \left( f_{\text{L}} - K x_1 \right) + \Delta_1(\mathbf{x}) \end{bmatrix}, \quad
\mathbf{g(x)} = \begin{bmatrix} 0 \\ b + \Delta_2(u) \end{bmatrix}.
\)

Next, consider the following matrix as the controllability matrix:
\begin{equation} \label{controllability matrix}
    M = \left[ \mathbf{g(x)} \quad \left[ \mathbf{f(x)}, \mathbf{g(x)} \right] \right],
\end{equation}
where \( \left[ \mathbf{f(x)}, \mathbf{g(x)} \right] = \frac{\partial \mathbf{g(x)}}{\partial x} \mathbf{f(x)} - \frac{\partial \mathbf{f(x)}}{\partial x} \mathbf{g(x)} \). Therefore,
\begin{equation}
    \left[ \mathbf{f(x)}, \mathbf{g(x)} \right] = \left[ \begin{array}{c} b + \Delta_2(u) \\ \left( b + \Delta_2(u) \right) \left( \frac{1}{m} \left( \frac{\partial f_{\text{L}}}{\partial x_2} + \frac{\partial \Delta_1(\mathbf{x})}{\partial x_2} \right) \right) \end{array} \right].
\end{equation}

Substituting this result into Eq.~\ref{controllability matrix}, the following expression is obtained:
\begin{equation}
    M = \left[ \begin{array}{cc} 0 & b + \Delta_2(u) \\ b + \Delta_2(u) & \left( b + \Delta_2(u) \right) \left( \frac{1}{m} \left( \frac{\partial f_{\text{L}}}{\partial x_2} + \frac{\partial \Delta_1(\mathbf{x})}{\partial x_2} \right) \right) \end{array} \right].
\end{equation}

Finally, by calculating the rank of the controllability matrix \( M \), the controllability of the system can be established. If \( \det(M) \neq 0 \), then \( M \) is full rank. Therefore,
\begin{equation}
    \det(M) =  -(b + \Delta_2(u))^2.
\end{equation}
Since \( b = \frac{1}{m} \neq 0 \) and \( \Delta_2(u) \neq -b \), the controllability of the system is confirmed as \( M \) is full rank. In other words, if \( \Delta_2(u) = -b \), it implies the absence of an actuator, meaning no control command can be applied to the system. Naturally, without any actuator or control commands, the concept of controllability becomes meaningless.

\subsection{Simple Learning Method}
\label{subsec: Simple Learning}

By defining the desired trajectory as $y_d$, with the tracking error given by:
\begin{equation} \label{error}
    e = y - y_d,
\end{equation}
Since this is a second-order system as defined in Sec.~\ref{subsec: dynamical system}, a sliding surface must be defined to account for velocity error as well. The sliding surface is defined as:
\begin{equation}
    s = \dot{e} + \lambda e.
\end{equation}

Considering the Lyapunov function as:
\begin{equation}
     V = \frac{1}{2}\left( s^2 + \frac{1}{\Gamma} \tilde{W}^T \tilde{W} \right),
\end{equation}
where $\Gamma$ is a positive constant, and $\tilde{W} = \hat{W} - W$. Since $W$ is a constant matrix of fixed coefficients, $\dot{\tilde{W}} = \dot{\hat{W}}$, which corresponds to the Neural Network updating rule. Taking the derivative of this Lyapunov function, the following expression is obtained:
\begin{equation}
     \begin{array}{l}\dot{V} = s\dot{s} + \frac{1}{\Gamma} \tilde{W}^T \dot{\hat{W}} \\ = s\left( \ddot{e} + \lambda \dot{e} \right) + \frac{1}{\Gamma} \tilde{W}^T \dot{\hat{W}} \\ = s\left( f + b u + \Delta - \ddot{y}_d + \lambda \dot{e} \right) + \frac{1}{\Gamma} \tilde{W}^T \dot{\hat{W}}, \end{array}
\end{equation}
if the updating rule is defined as follows:
\begin{equation} \label{updating rule}
     \dot{\hat{W}} = \Gamma \mu s,
\end{equation}
the resulting expression is:
\begin{equation} \label{lyapunov before u}
     \dot{V} = s\left( f + b u + \Delta - \ddot{y}_d + \lambda \dot{e} \right) + \tilde{W}^T \mu s.
\end{equation}

The control command can now be expressed as:
\begin{equation} \label{control command}
     u = b^{-1} \left( -f - \hat{\Delta} + \ddot{y}_d - k_\text{C} s - \lambda \dot{e} \right),
\end{equation}
where $k_\text{C}$ is a positive constant.

By substituting Eq.\ref{control command} into Eq.\ref{lyapunov before u}, the expression becomes:
\begin{equation}
     \dot{V} = s\left( \Delta - \hat{\Delta} - k_\text{C} s \right) + \tilde{W}^T \mu s,
\end{equation}
since $\tilde{\Delta} = \hat{\Delta} - \Delta$, and thus $ \tilde{W}^T \mu = \tilde{\Delta} + \varepsilon$, this leads to the following expression:
\begin{equation}
    \dot{V} = s\left( -\tilde{\Delta} - k_\text{C} s \right) + s\left( \tilde{\Delta} + \varepsilon \right) = s\left( -k_\text{C} s + \varepsilon \right).
\end{equation}

This shows that $\dot{V} < 0$ when $\left\| k_\text{C} s \right\| \geq \left\| \varepsilon \right\|$, ensuring a bounded tracking error \cite{Khalil2015}. Determining optimal design parameters is typically done through trial and error.

\subsection{Composite Learning Method}
\label{subsec: Composite Learning}

Recent literature has introduced several modified learning approaches that significantly enhance the training of Neural Network parameters. One such development is the composite learning method, which integrates multiple learning strategies to optimize performance. Various composite learning approaches have been proposed, all of which incorporate estimation performance into the updating law \cite{Xu2014, Xu2017, Xu2021}. These methods result in faster learning and improved precision \cite{Xu2019}. Specifically, state estimation can be formulated as:
\begin{equation}
     \ddot {\hat y} = f + bu + \hat \Delta - k_\text{SE} s_\text{D} - \lambda_\text{D} \dot e_\text{D},
\end{equation}
where $e_\text{D} = \hat y - y$ is State Estimation error, $s_\text{D} = \dot {e_\text{D}} + \lambda_\text{D} e_\text{D}$ is State Estimation sliding surface, and $k_\text{SE}$ and $\lambda_\text{D}$ are positive constants. By considering the following Lyapunov function:
\begin{equation} \label{lyapunov composite}
     V = \frac{1}{2}\left( {s^2 + k_\text{D} s_\text{D}^2 + \frac{1}{\Gamma }{{\tilde W}^T}\tilde W} \right).
\end{equation}

The control command remains unchanged as in Eq.~\ref{control command}, by taking the derivative of this Lyapunov function, the expression can be written as:
\begin{equation}
    \begin{array}{l}
        \dot{V} = s \dot{s} + k_\text{D} s_\text{D} \dot{s}_\text{D} + \frac{1}{\Gamma} \tilde{W}^T \dot{\hat{W}} \\ 
        = s\left( \ddot{e} + \lambda \dot{e} \right) + k_\text{D} s_\text{D} \left( \ddot{e}_\text{D} + \lambda_\text{D} \dot{e}_\text{D} \right) + \frac{1}{\Gamma} \tilde{W}^T \dot{\hat{W}} \\ 
        = s\left( \ddot{y} - \ddot{y}_d + \lambda \dot{e} \right) + k_\text{D} s_\text{D} \left( \ddot{\hat{y}} - \ddot{y} + \lambda_\text{D} \dot{e}_\text{D} \right) + \frac{1}{\Gamma} \tilde{W}^T \dot{\hat{W}},
    \end{array}
\end{equation}
if the updating rule in Eq.~\ref{updating rule} modified as:
\begin{equation} \label{updating rule composite}
    \dot {\hat W} = \Gamma \mu \left( {s - k_\text{D} s_\text{D}} \right),
\end{equation}
by substituting Eq.~\ref{control command}, the following result is obtained:
\begin{equation}
    \dot{V} = s\left( -k_\text{C} s + \varepsilon \right) + k_\text{D} s_\text{D} \left( -k_\text{SE} s_\text{D} - \varepsilon \right).
\end{equation}

As well as Simple Learning this shows that $\dot{V} < 0$ when $\left\| \frac{{k_\text{C} s^2 + k_\text{D} k_\text{SE} s_\text{D}^2}}{{s - k_\text{D} s_\text{D}}} \right\| \ge \left\| \varepsilon \right\|$, ensuring a bounded tracking error. The controller's performance is discussed and evaluated in Sec.~\ref{subsec: eval controller}.

\section{Results and Discussion}
\label{sec: results}

In this section, the simulation results of the 2DOF cylinder system with a Neural Network-based controller are presented. First, the fixed and free vibrations of the model are analyzed. Next, the performance of the simulated model is validated through a comparison with the literature. Following this, the effectiveness of the Neural Network in uncertainty estimation is investigated. Finally, the performance of the Neural Network-based controller in reducing the cylinder's vibration amplitude, along with the corresponding control commands, is presented.
Table \ref{tab_parameters} presents the cylinder model parameters.

\begin{table}[H]
\small
\centering
\caption{Parameters of the Cylinder Model}
\label{tab_parameters}
\begin{tabularx}{\textwidth}{l X X X}
    \toprule
    Description & Parameter & Value & Unit \\
    \midrule
    Cylinder Mass & $m$ & 1.571 & kg \\
    Cylinder Diameter & D & 1 & m \\
    Flow Kinematic Viscosity & $\nu$ & 0.00667 & m$^2$/s \\
    Flow Velocity & $U$ & 1 & m/s \\
    Spring Stiffness & K & 2.48 & N/m \\
    Reduced Velocity & $U_r$ & 5 & - \\
    Reynolds Number & $Re$ & 150 & - \\
    \bottomrule
\end{tabularx}
\end{table}

\subsection{Fixed \& Free Vibration}
\label{subsec: fixed & free}

\begin{figure}[H]
     \centering
     
     \begin{subfigure}[b]{0.05\textwidth}
         \centering
         \includegraphics[width=\textwidth]{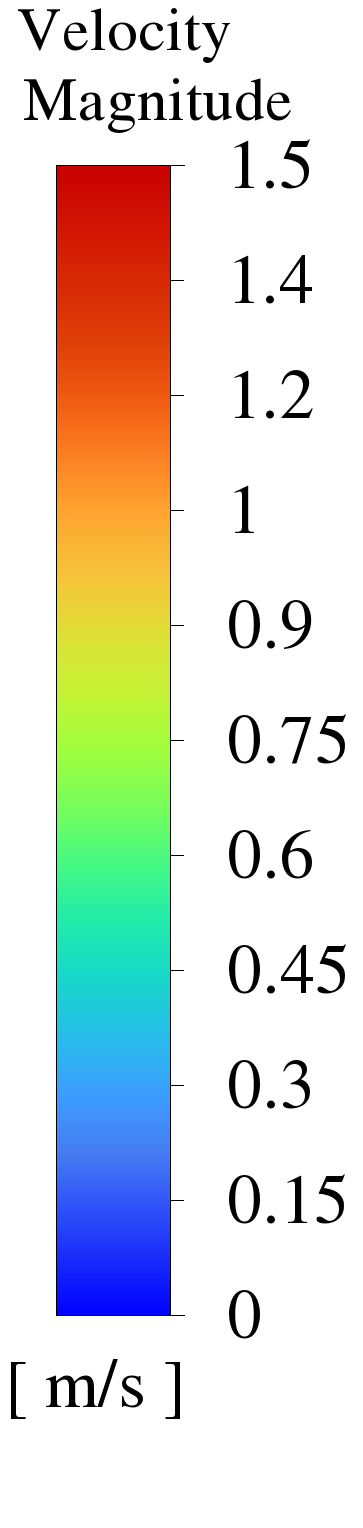}
        \caption*{}
     \end{subfigure}
     \hfill
     \begin{subfigure}[b]{0.46\textwidth}
         \centering
         \includegraphics[width=\textwidth]{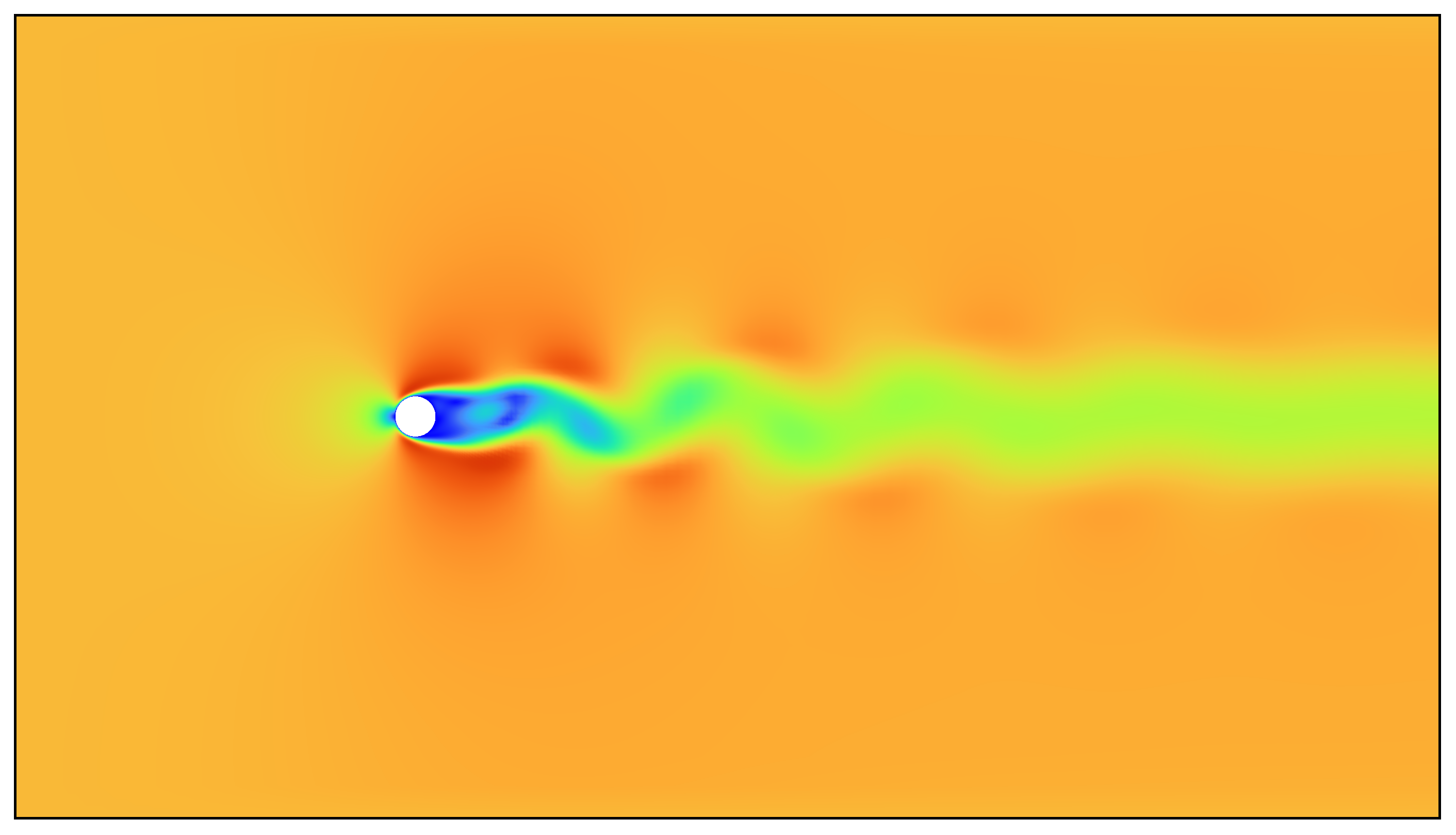}
         \caption{Velocity magnitude - Fixed case}
     \end{subfigure}
     \hfill
     \begin{subfigure}[b]{0.46\textwidth}
         \centering
         \includegraphics[width=\textwidth]{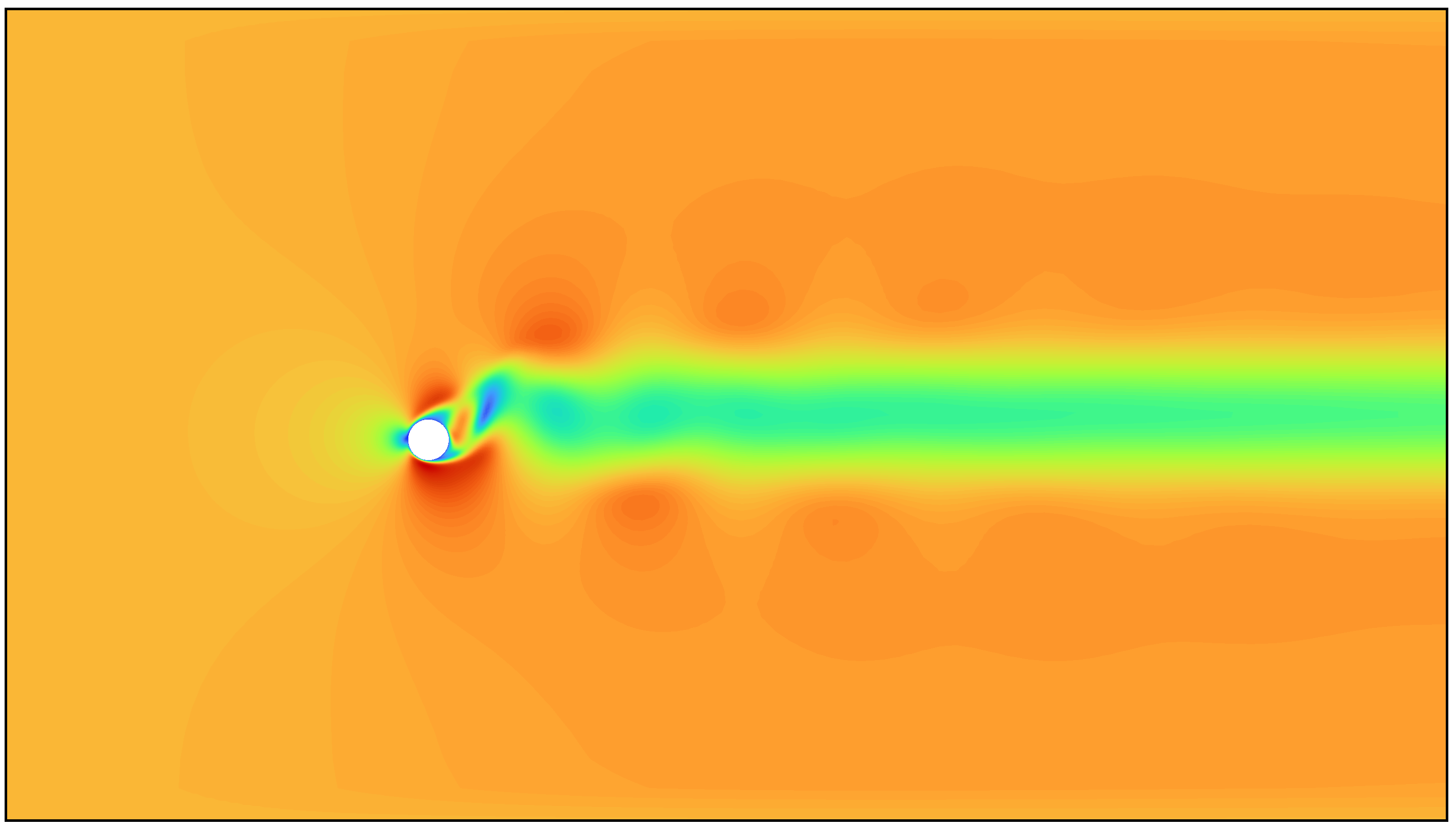}
         \caption{Velocity magnitude - Free case}
     \end{subfigure}

     \begin{subfigure}[b]{0.05\textwidth}
         \centering
         \includegraphics[width=\textwidth]{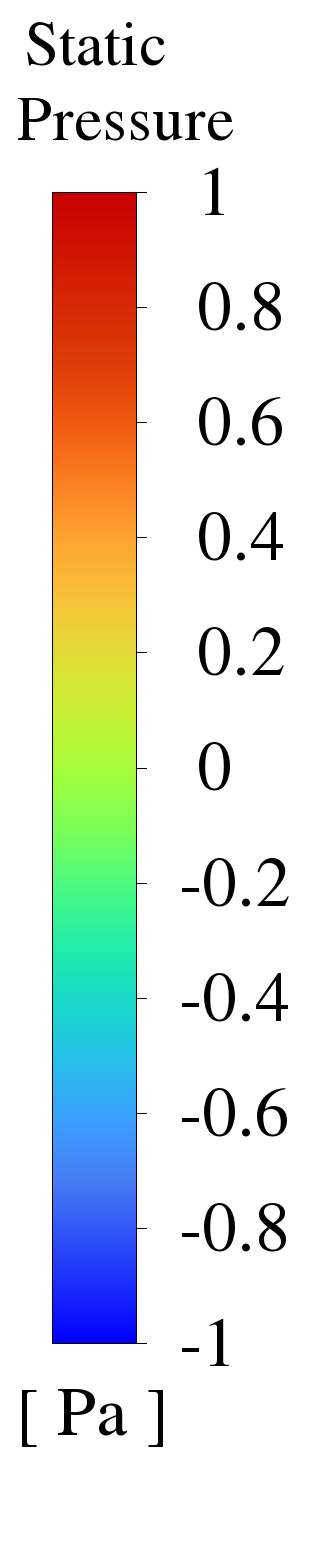}
         \caption*{}
     \end{subfigure}
     \hfill
     \begin{subfigure}[b]{0.46\textwidth}
         \centering
         \includegraphics[width=\textwidth]{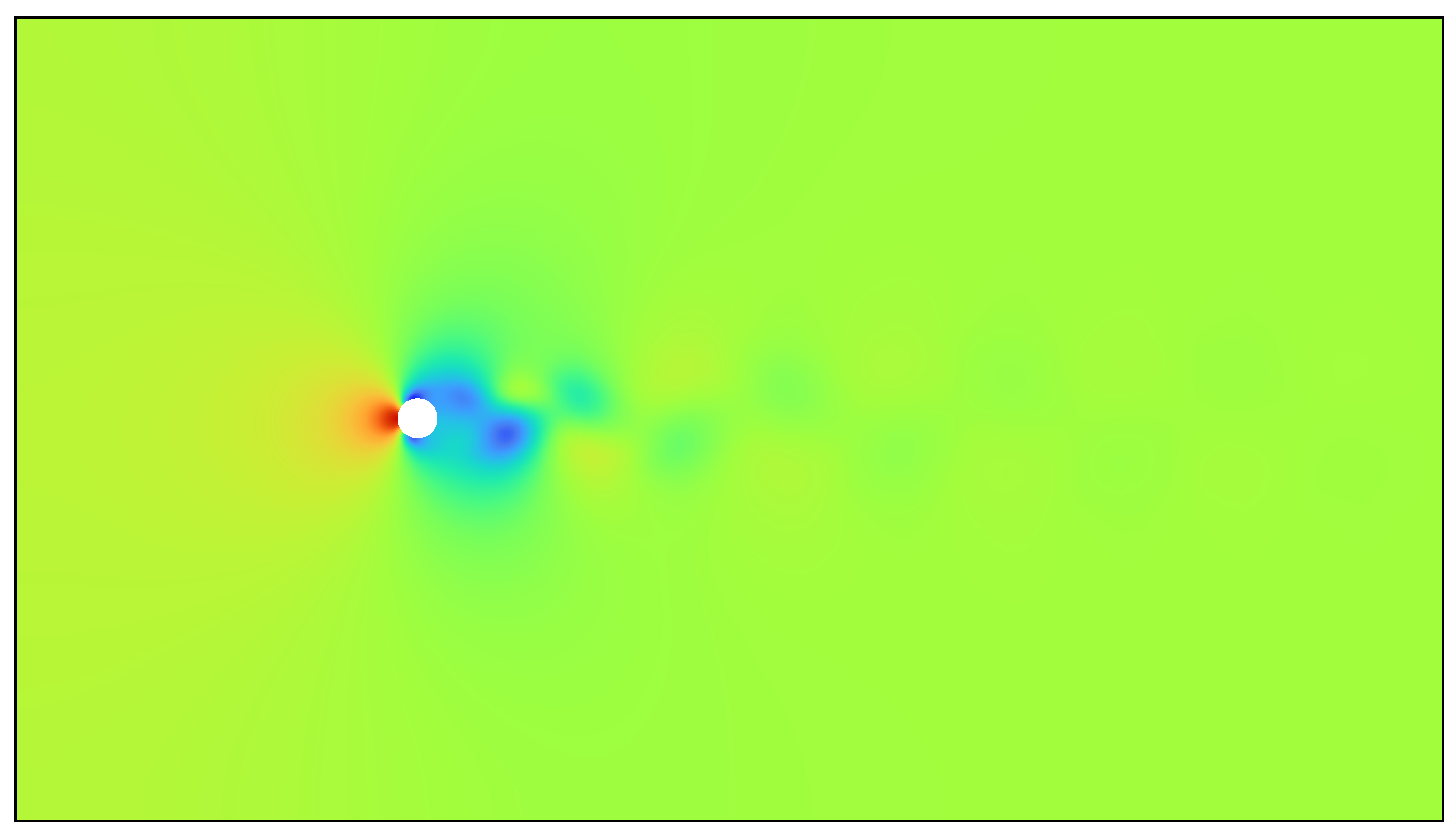}
         \caption{Static pressure - Fixed case}
     \end{subfigure}
          \hfill
     \begin{subfigure}[b]{0.46\textwidth}
         \centering
         \includegraphics[width=\textwidth]{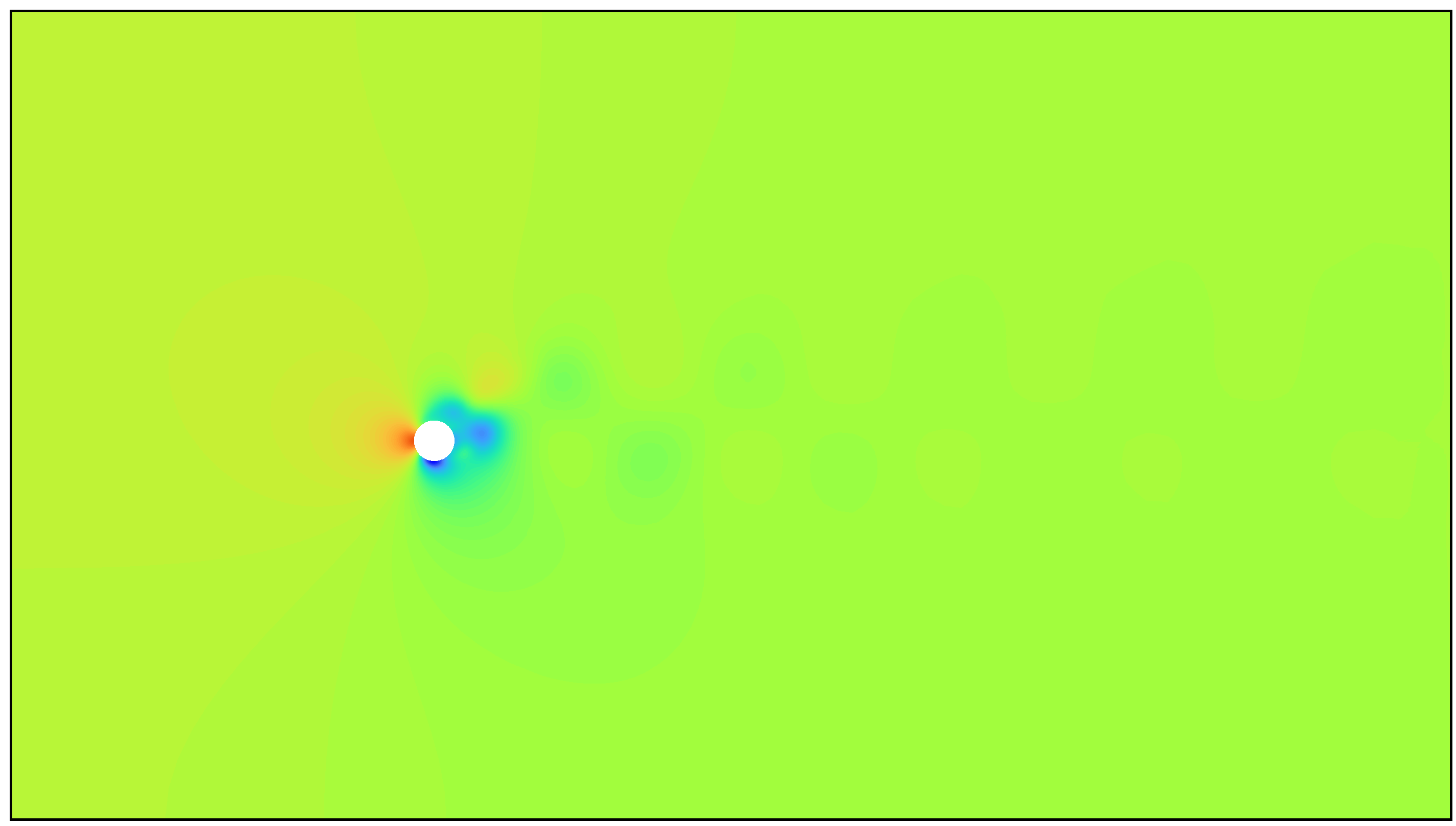}
         \caption{Static pressure - Free case}
     \end{subfigure}

    \begin{subfigure}[b]{0.05\textwidth}
         \centering
         \includegraphics[width=\textwidth]{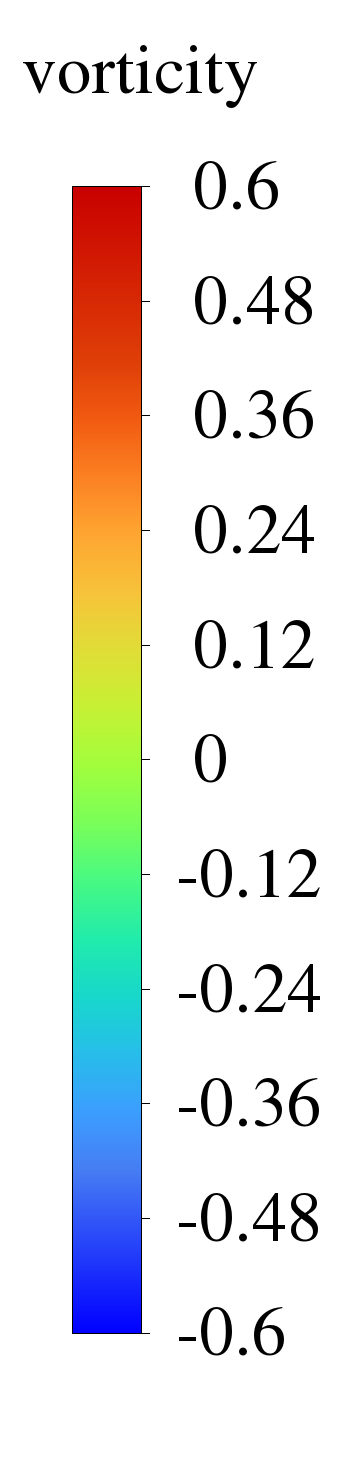}
         \caption*{}
     \end{subfigure}
     \hfill
     \begin{subfigure}[b]{0.46\textwidth}
         \centering
         \includegraphics[width=\textwidth]{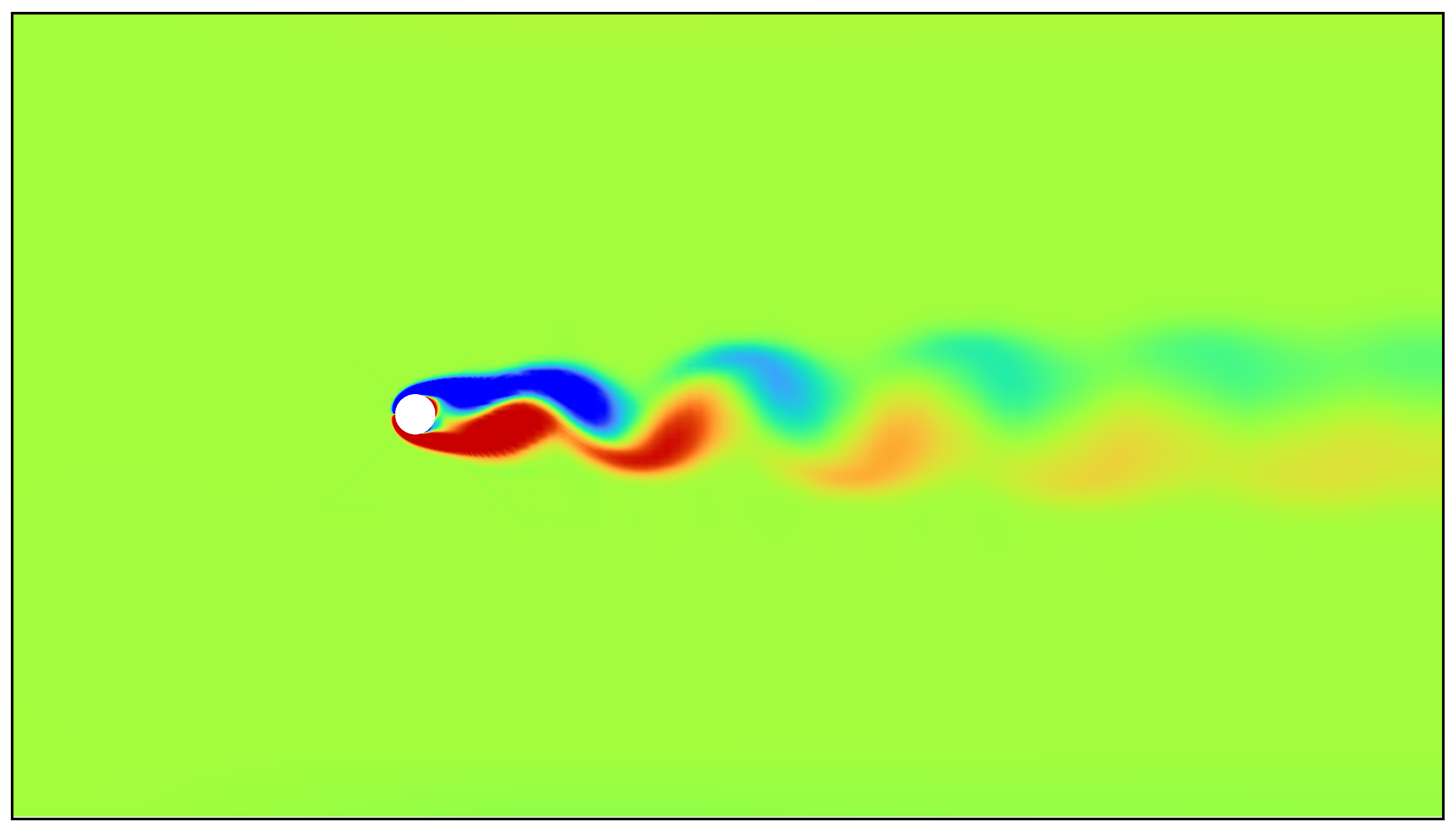}
         \caption{Vorticity magnitude - Fixed case}
     \end{subfigure}
     \hfill
     \begin{subfigure}[b]{0.46\textwidth}
         \centering
         \includegraphics[width=\textwidth]{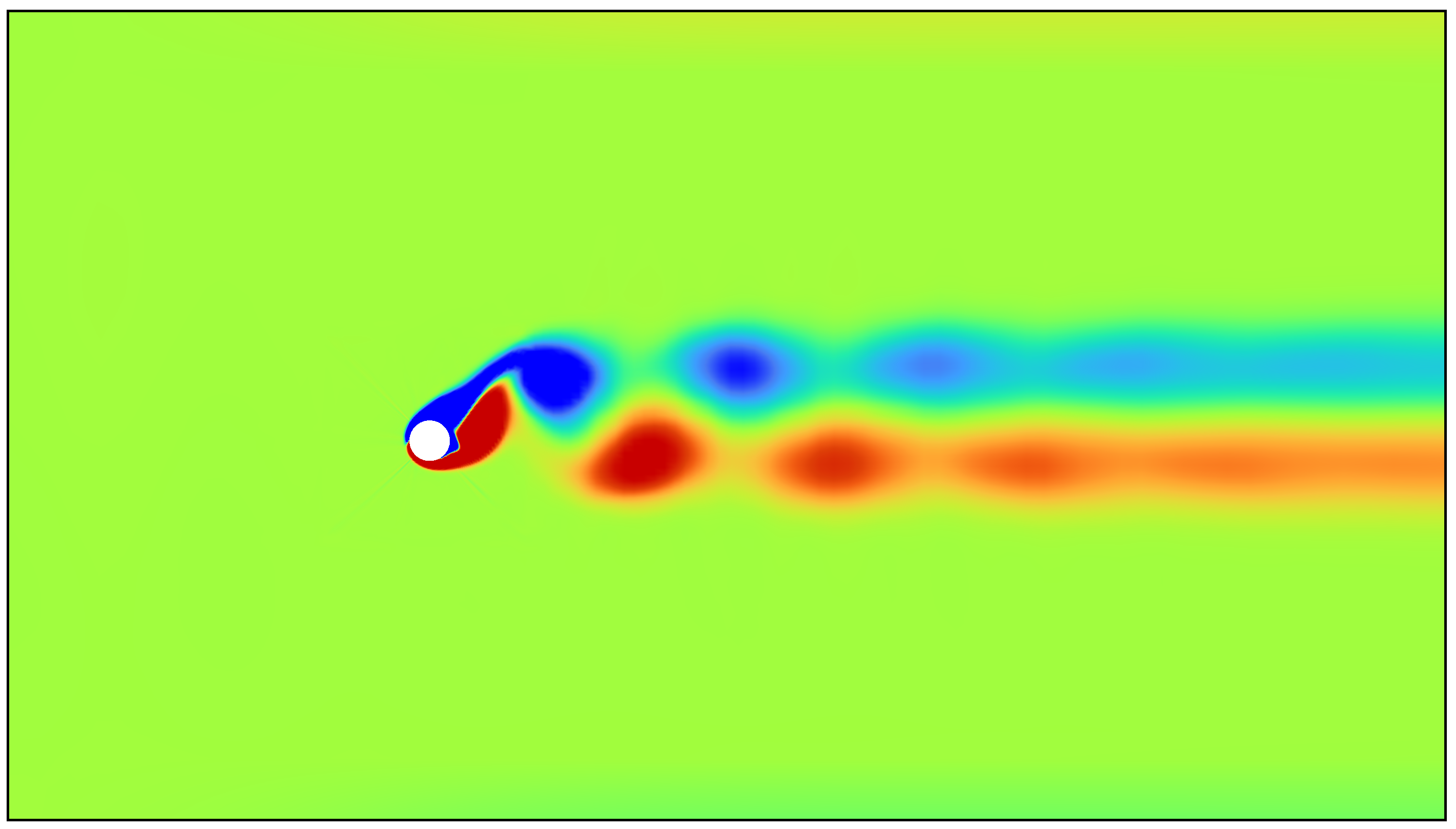}
         \caption{Vorticity magnitude - Free case}
     \end{subfigure}

        \caption{Comparison of fluid domain results for both fixed and free case}
        \label{fixedcont}

\end{figure}

In this section, the results of simulations involving the physics of vortices and cylinder vibrations are presented for two cases with a reduced velocity of $U_r=5$ in the absence of any uncertainty. In the first case, the cylinder is fixed, with no movement in the $x$ or $y$ directions. The second case allows free movement of the cylinder in both the $x$ and $y$ directions for comparison. Fluid domain parameters such as velocity, pressure, and vorticity are compared for both the fixed and free cases in Fig.~\ref{fixedcont}, illustrating that due to the large vertical displacement, the generated vortices are larger in size and magnitude and more spaced out.

\begin{figure}[H]
     \centering
     \begin{subfigure}[b]{0.49\textwidth}
         \centering
         \includegraphics[width=\textwidth]{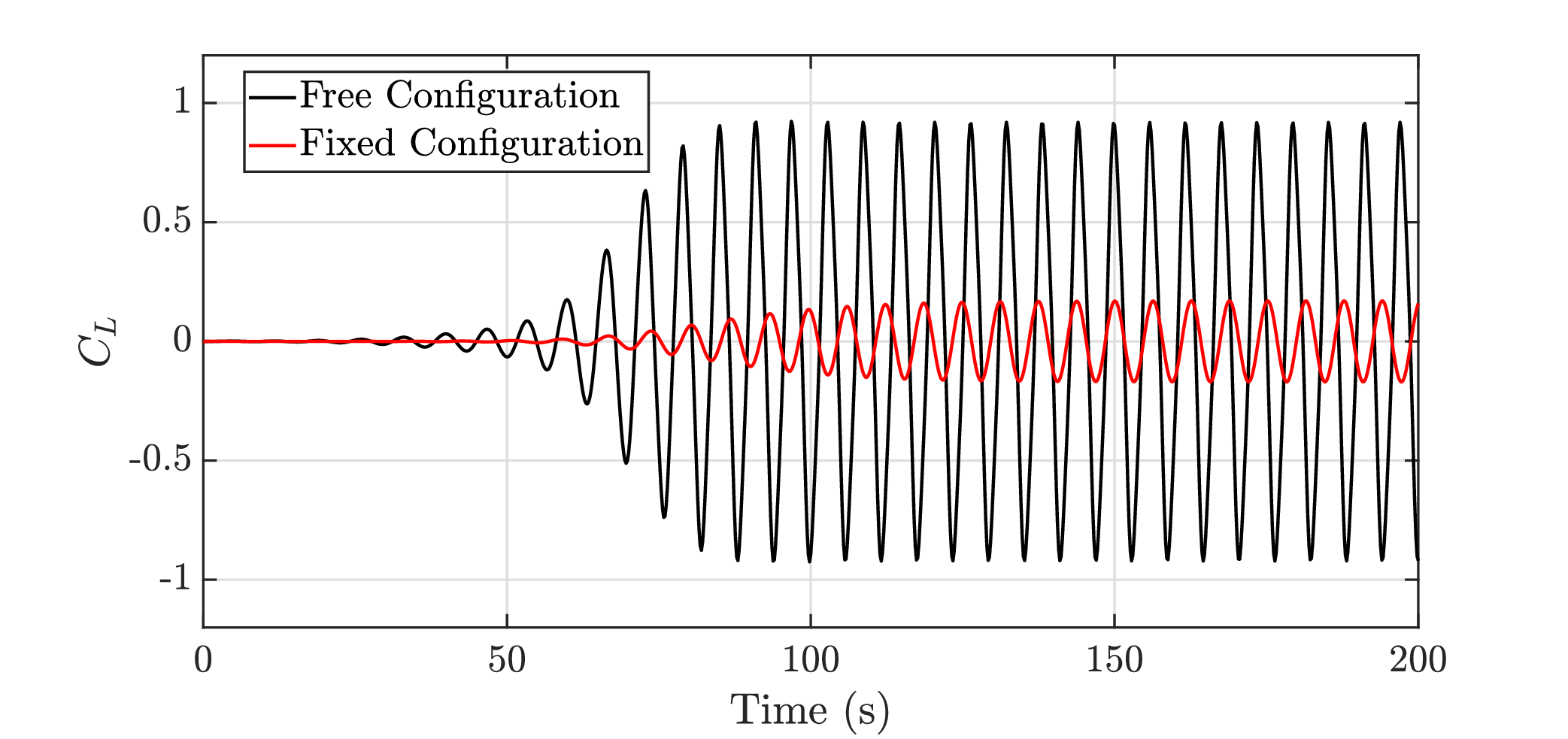}
         \caption{Lift coefficient}
     \end{subfigure}
     \hfill
     \begin{subfigure}[b]{0.49\textwidth}
         \centering
         \includegraphics[width=\textwidth]{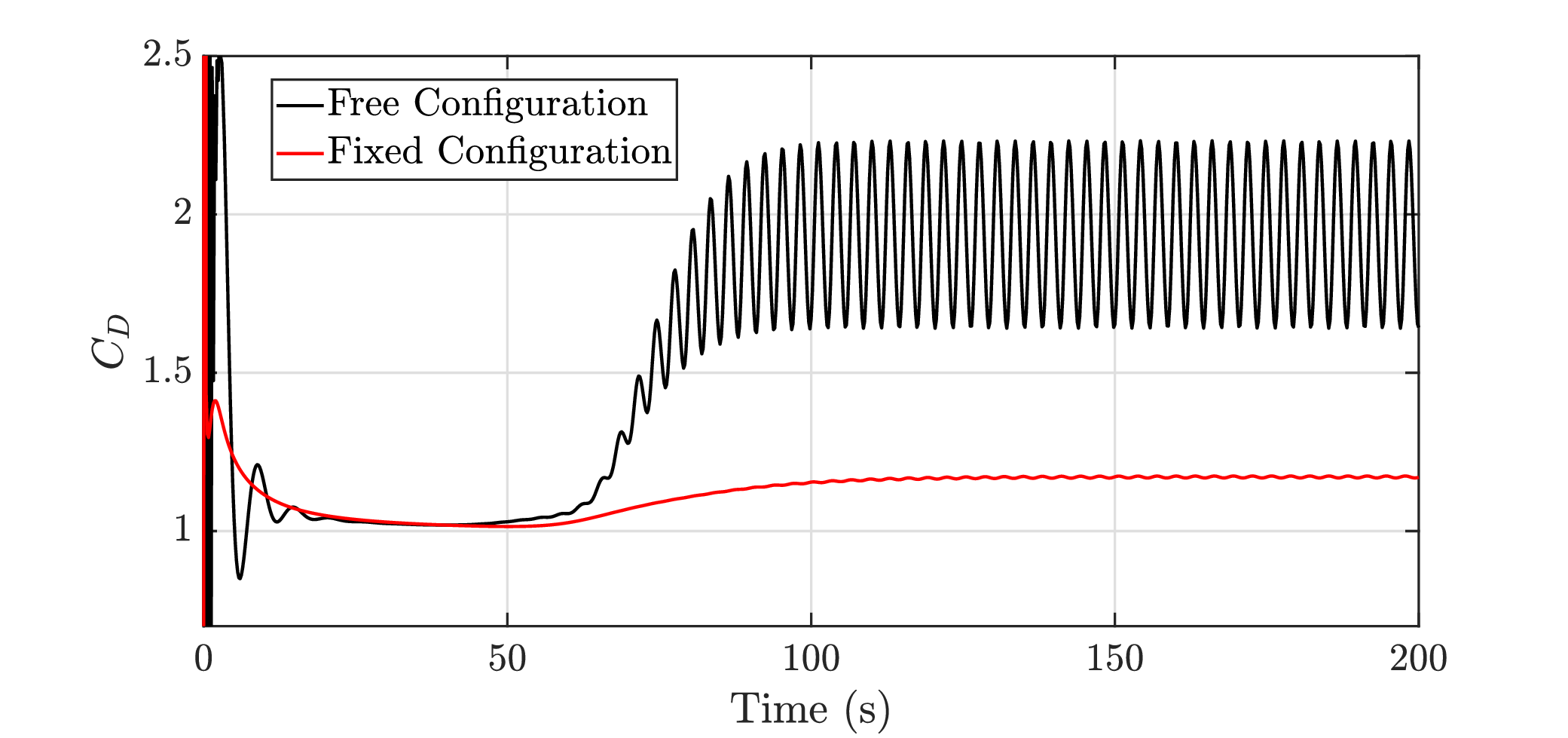}
         \caption{Drag coefficient}
     \end{subfigure}
        \caption{Lift and Drag coefficient for fixed and free case}
        \label{fixvsfree}
\end{figure}

Fig.~\ref{fixvsfree} presents the variation of lift and drag coefficients over time for both the fixed cylinder and the free vibration case. In the fixed cylinder scenario, the drag coefficient shows a relatively steady pattern with small oscillations, while the lift coefficient exhibits more significant fluctuations due to the periodic shedding of vortices in the wake. The average drag coefficient ($C_D$) and maximum lift coefficient ($C_L$) were found to be 1.15 and 0.23, respectively, and these results are compatible with those reported in \cite{modarres2022flow}. In contrast, the free vibration case demonstrates more complex behavior, with both lift and drag coefficients varying significantly over time. The amplitude of oscillations in the free vibration case is noticeably larger, with higher frequencies compared to the fixed cylinder, indicating more dynamic flow interaction.
Additionally, the mean values of drag coefficient in the free vibration case is  increased, reflecting the enhanced fluid-structure interaction due to the cylinder’s freedom of movement in the $x$ and $y$ directions.

\begin{figure}[H]
    \centering
    \begin{subfigure}[b]{0.49\textwidth}
        \centering
        \includegraphics[width=\textwidth]{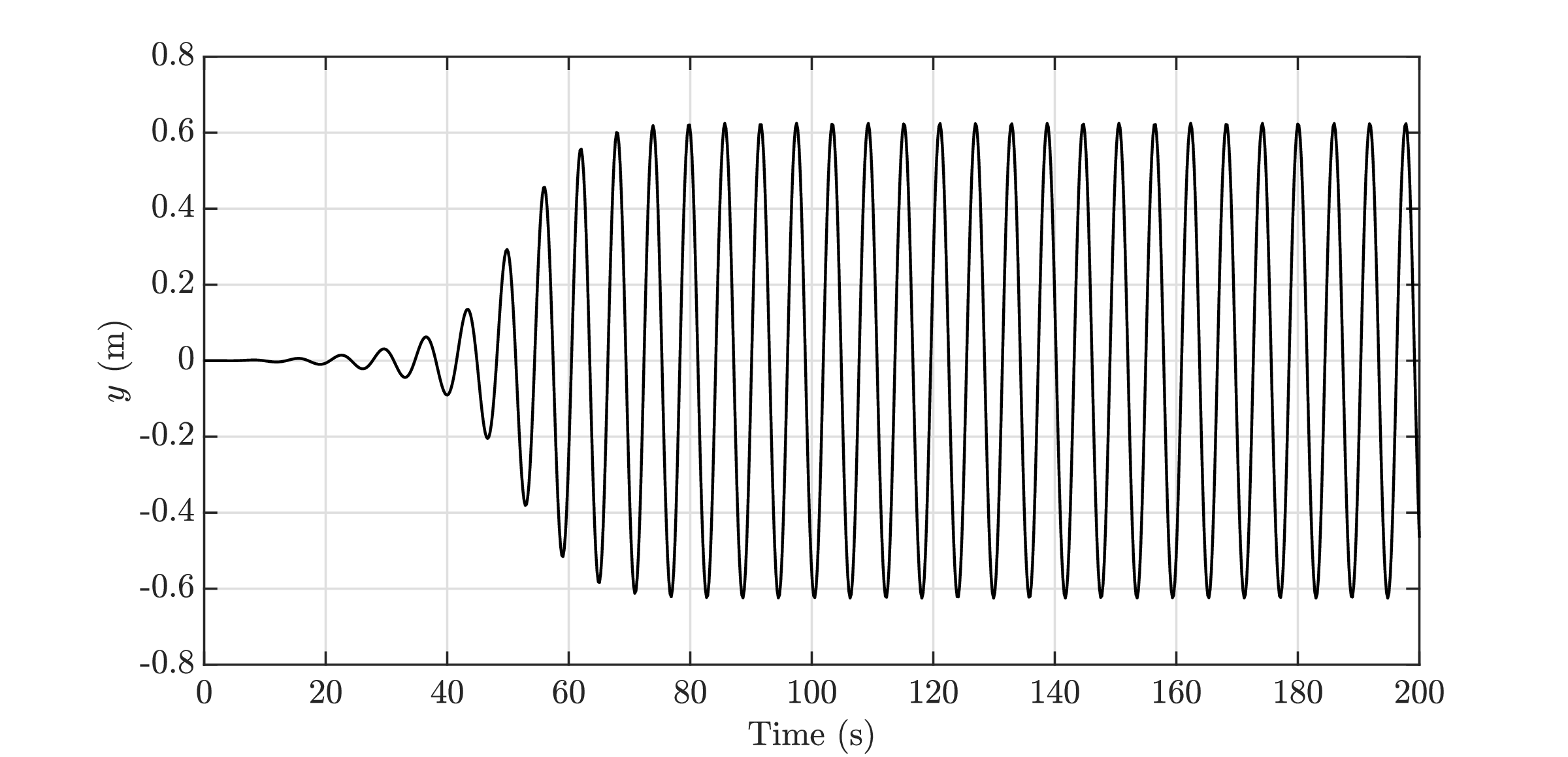}
        \label{figa_freevib}
        \caption{$y$ position displacement}
    \end{subfigure}%
    \begin{subfigure}[b]{0.49\textwidth}
        \centering
        \includegraphics[width=\textwidth]{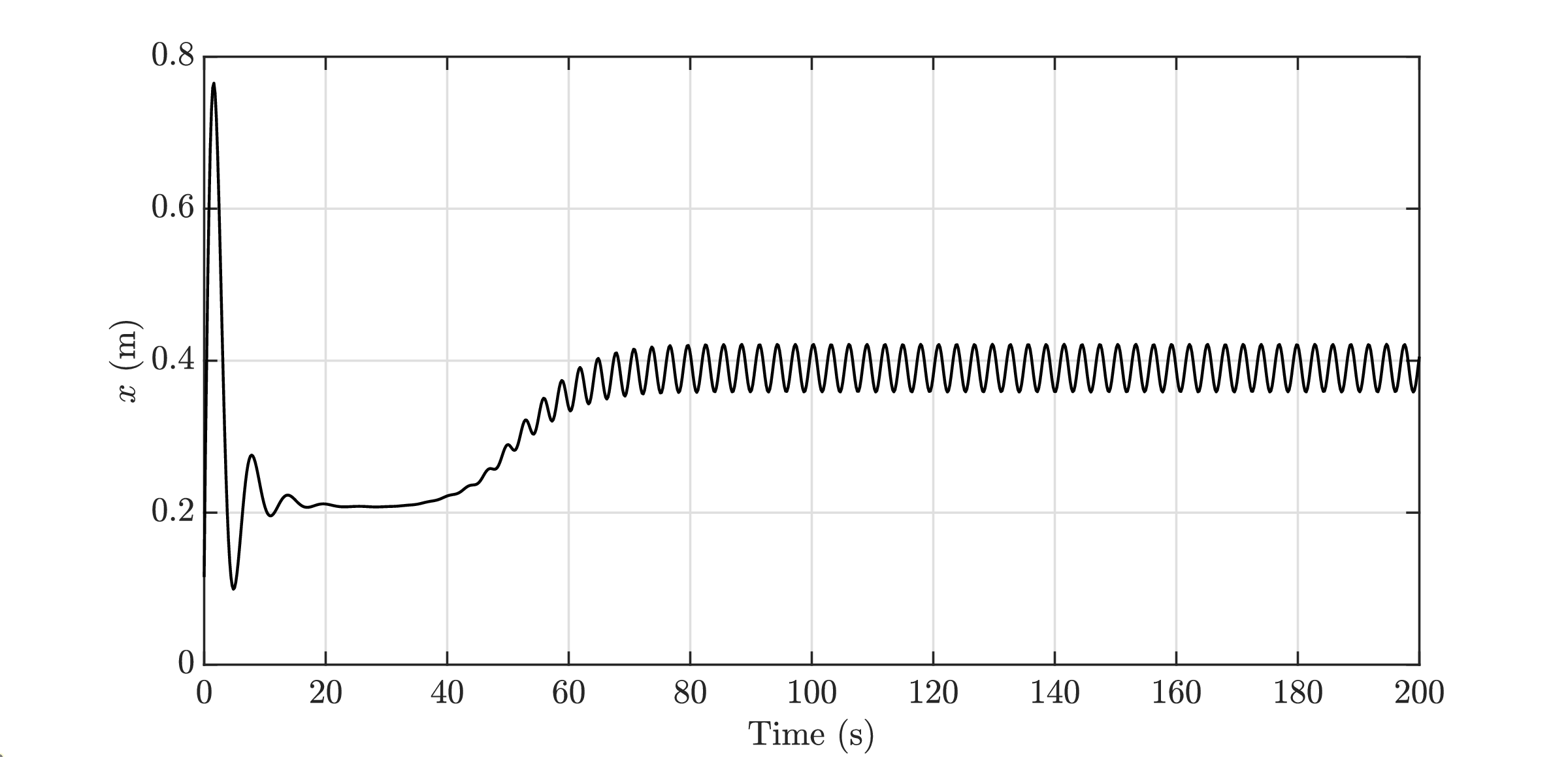}
        \label{figb_freevib}
        \caption{$x$ position displacement}
    \end{subfigure}%
    \caption{Free Vibration in $y$ and $x$ directions in the absence of uncertainty}
    \label{free vibration}
\end{figure}

Figure \ref{free vibration} illustrates the free vibration behavior of the cylinder in the absence of uncertainties, capturing the influence of lift and drag force patterns on the motion in both the vertical ($y$) and horizontal ($x$) directions. The y-direction oscillations are more pronounced due to the dominant effect of lift forces, while the x-direction motion, driven by drag forces, is relatively subdued. Over time, the system stabilizes, with the y-direction oscillations reaching a steady level after approximately 80 seconds.  This behavior provides a baseline understanding of the system's dynamic response with no uncertainty.

\begin{figure}[H]
    \centering
    \begin{subfigure}[b]{0.48\textwidth}
        \centering
        \includegraphics[width=\textwidth]{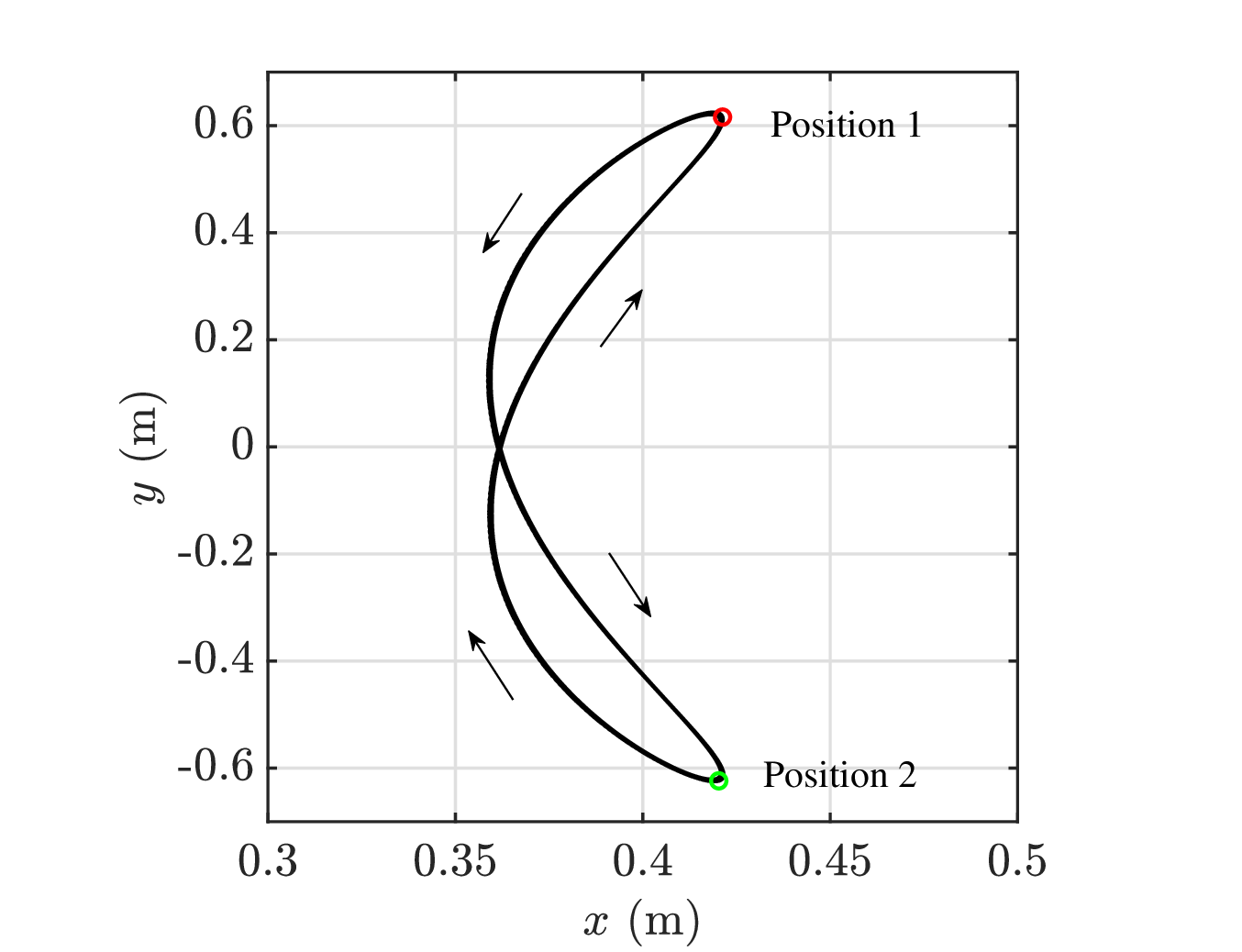}
        \label{figa_fr}
        \caption{Cylinder motion trajectory}
    \end{subfigure}
    \begin{subfigure}[b]{0.47\textwidth}
        \centering
        \includegraphics[width=\textwidth]{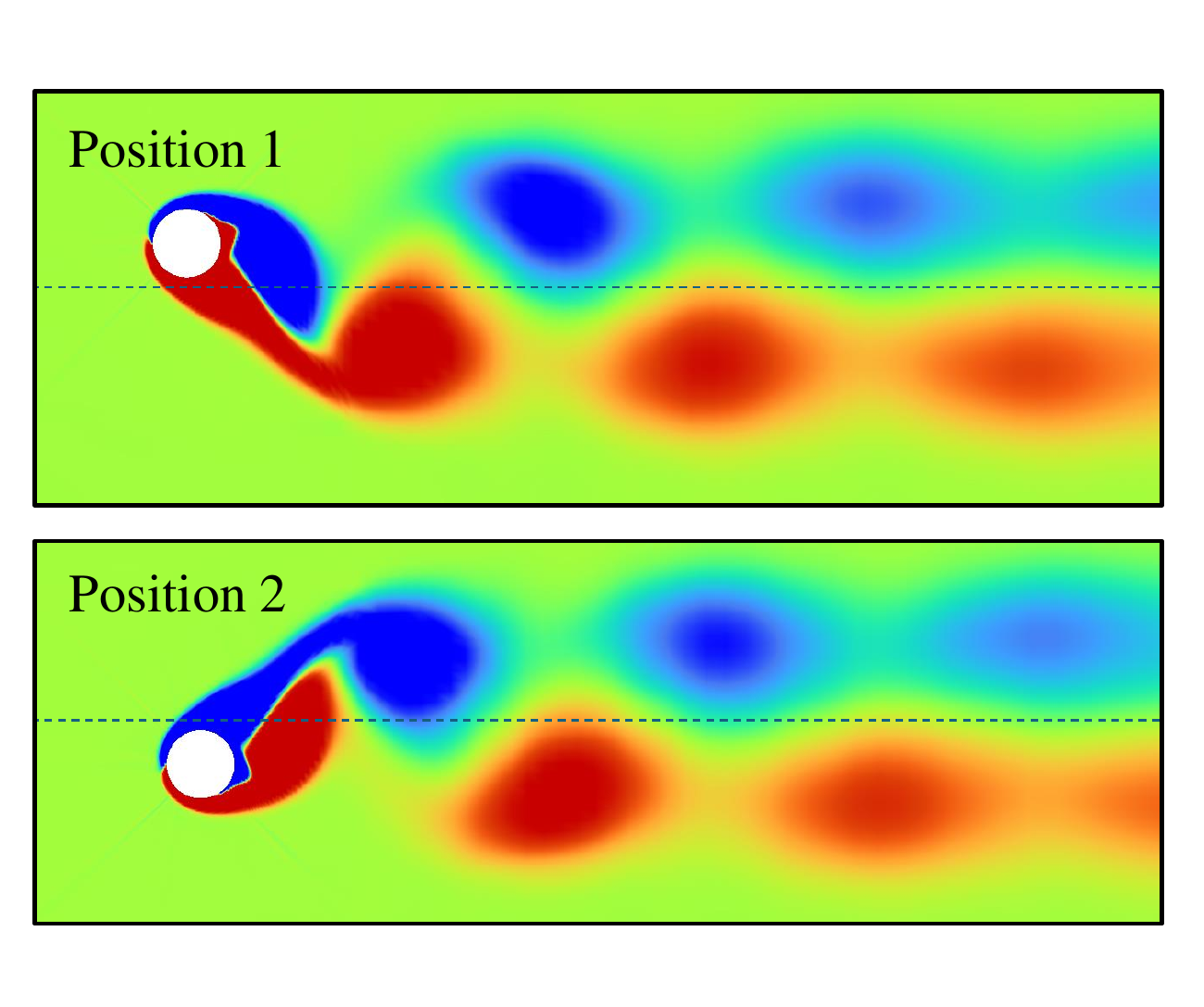}
        \label{figb_freevibvv}
        \caption{Vorticity magnitude contour }
    \end{subfigure}
    \caption{Orbital motion trajectory and vorticity  in the highest and lowest positions}
    \label{free_vibratiovvv}
\end{figure}

Figure \ref{free_vibratiovvv} provides a detailed visualization of the cylinder’s trajectory in the XY plane and the corresponding vorticity magnitudes at its highest and lowest positions. The trajectory plot is generated after the cylinder reaches a steady mean oscillation value, confirming the stability of the motion. The trajectory plot reveals a figure-eight-shaped orbital path, which arises due to the fluctuating drag force exciting the motion at a frequency twice that of the lift force \cite{NEPALI2020102991}. This dynamic interplay between the two forces results in the distinctive trajectory pattern. Furthermore, the cylinder exhibits both clockwise and counterclockwise movement, depending on whether it is in the upper or lower half of its oscillation cycle.

\subsection{Validation of the FSI model}
\label{subsec: eval fluid model}

In this section, the validity of the numerical model for FSI simulation has been examined through a comparative analysis with similar works. In the validation case, the free vibration of a cylinder without the implementation of any control mechanism has been compared with two notable previous numerical studies.
In first study \citet{bao2012two} conducted a comprehensive investigation involving 2D numerical models of vibrating cylinders, exploring various isolated and tandem configurations across a range of natural frequencies. 
In next one, \citet{verma2022three} performed a 3D simulation of a circular cylinder at different Reynolds numbers. Notably, the comparison between 2D and 3D simulations revealed minimal discrepancies in results at low Reynolds numbers.

\begin{figure}[H]
  \centering
  \includegraphics[width=0.8\linewidth]{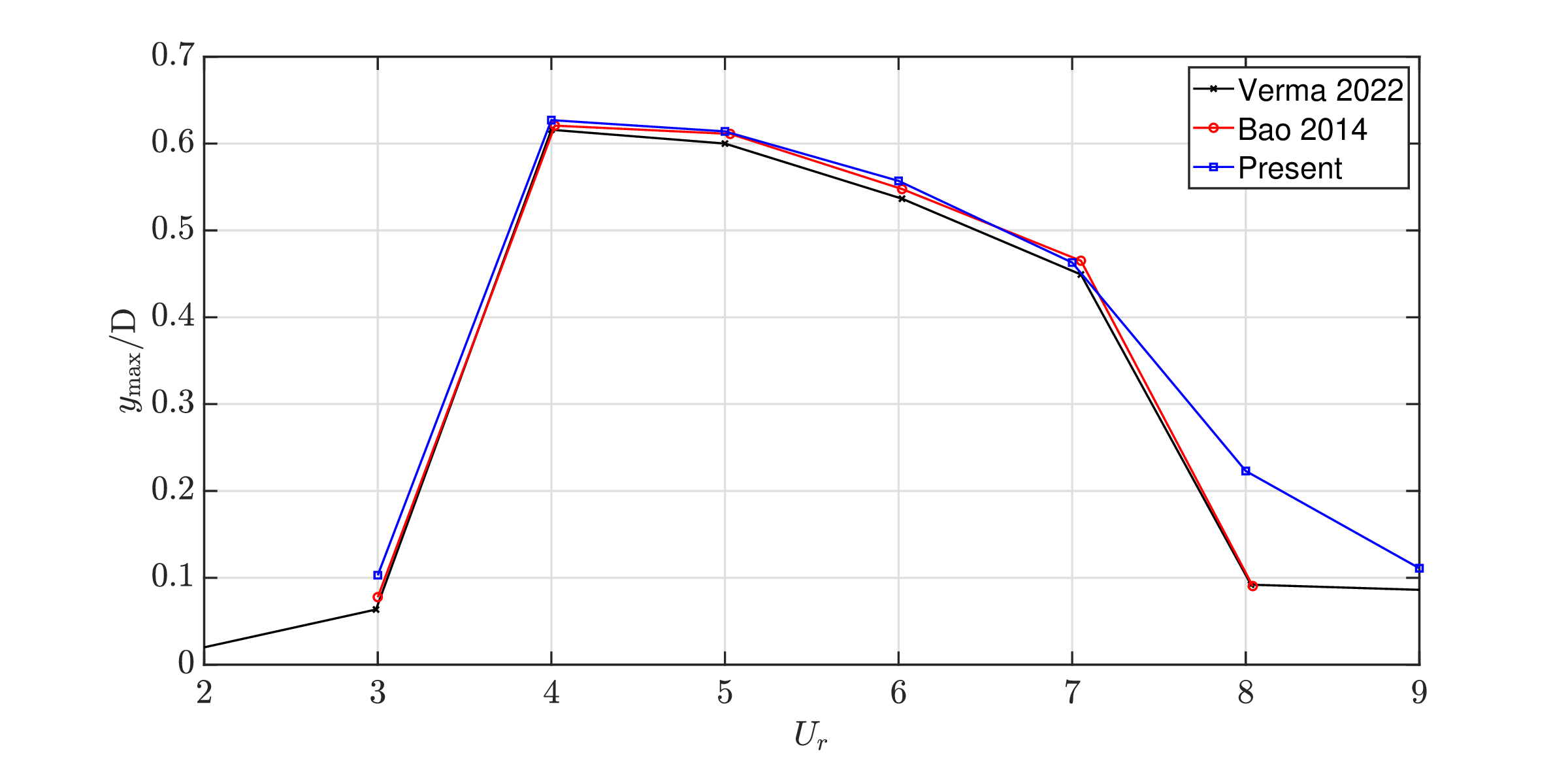}
  \caption{Validation results for maximum amplitude of traversed vibration}
  \label{validation}
\end{figure}

To maintain consistency and comparability with aforementioned studies, specific system parameters were established. These parameters include a Reynolds number of 150, operating within the laminar regime, and $m^*$ (Non-dimensional mass ratio) of 2.56, with no damping present, across different reduced velocities ranging from 3 to 9. The validation results for the maximum amplitude of traversed vibration at different reduced velocities are depicted in Fig.~\ref{validation}.

It is observable that the results align well with previous studies, particularly within the lock-in region ($5 < U_r < 7$), which is the focus of interest for controlled simulations.

To ensure comprehensive validation of a specific aspect within the lock-in region at $U_r = 5$, other parameters were analyzed and compared to those in the Verma study, as detailed in Table \ref{tab_Validation}. It is evident that the comparison demonstrates alignment between the results obtained from simulation and those from the Verma study.

To compare the dynamic and oscillation characteristics of the cylinder, following parameters (as presented in Table \ref{tab_Validation}) has been compared. $x'/\text{D}$ is the oscillating amplitude of in-line oscillation non-dimensionalized by the cylinder diameter, $y_{\text{max}}/\text{D}$ represents the maximum transverse oscillation non-dimensionalized in the same way, and $f_y/f_{ny}$ denotes the ratio of the oscillation frequency to the natural frequency of the system in the y-direction.
By presenting data of obtained results and comparing them with other previous similar works, the accuracy of cylinder vibration simulation has been confirmed.

\begin{table}[H]
\centering
\caption{Amplitude and Frequency Response for $Ur=5$}
\label{tab_Validation}
\small
\begin{tabularx}{\textwidth}{X X X X}
    \toprule
    & $x'/\text{D}$ & $y_{\text{max}}/\text{D}$ & $f_y/f_{ny}$ \\
    \midrule
    Verma \cite{verma2022three} & 0.027 & 0.608 & 0.944 \\
    Present Study & 0.032 & 0.615 & 0.919 \\
    Difference (\%) & 18.5 & 1.1 & -2.6 \\
    \bottomrule
\end{tabularx}
\end{table}

\subsection{Evaluation of the Neural Network}
\label{subsec: eval nn}

\begin{figure}[H]
  \centering
  \includegraphics[width=0.8\linewidth]{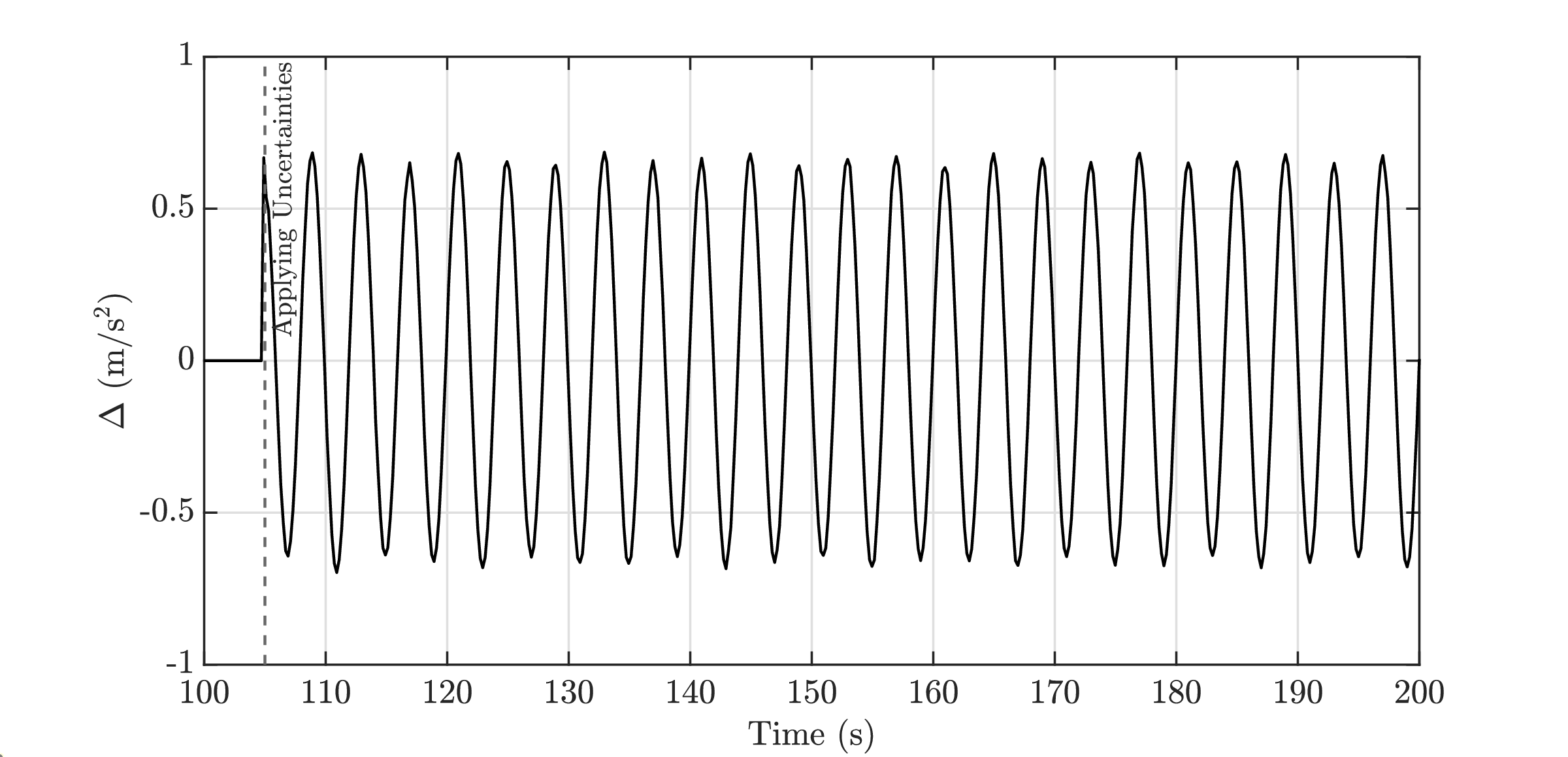}
  \caption{Actual uncertainty}
  \label{Delta}
\end{figure}

This section examines the performance of the Neural Network in estimating uncertainties. As defined in Sec.~\ref{subsec: dynamical system}, the model encounters an uncertainty term \( \Delta(y, \dot{y}, u) = \Delta_1(y, \dot{y}) + \Delta_2(u) u + \Delta_3 \). Table~\ref{tab_uncertainties} presents the values of the applied uncertainties for this study. The number of neurons in the hidden layer set to 15 ($n = 15$) for current study. Fig.~\ref{Delta} illustrates the uncertainty term (\( \Delta \)). As seen, \( \Delta \) is introduced starting at 105 seconds.

\begin{table}[H]
\centering
\caption{Values of Applied Uncertainties}
\label{tab_uncertainties}
\small
\begin{tabularx}{\textwidth}{l X X}
    \toprule
    Description & Parameter & Value \\
    \midrule
    20\% uncertainty in model parameters & $\Delta_1(y, \dot{y})$ & $-0.2\text{K}y$ \\
    20\% control command ineffectiveness (Actuator Fault) & $\Delta_2(u)$ & $-0.2b$ \\
    Constant, time-variant sine wave as an external disturbance & $\Delta_3$ & $0.5\sin{\frac{\pi t}{2}}$ \\
    \bottomrule
\end{tabularx}
\end{table}

\begin{figure}[H]
    \centering
    \begin{subfigure}[b]{0.49\textwidth}
        \centering
        \includegraphics[width=\textwidth]{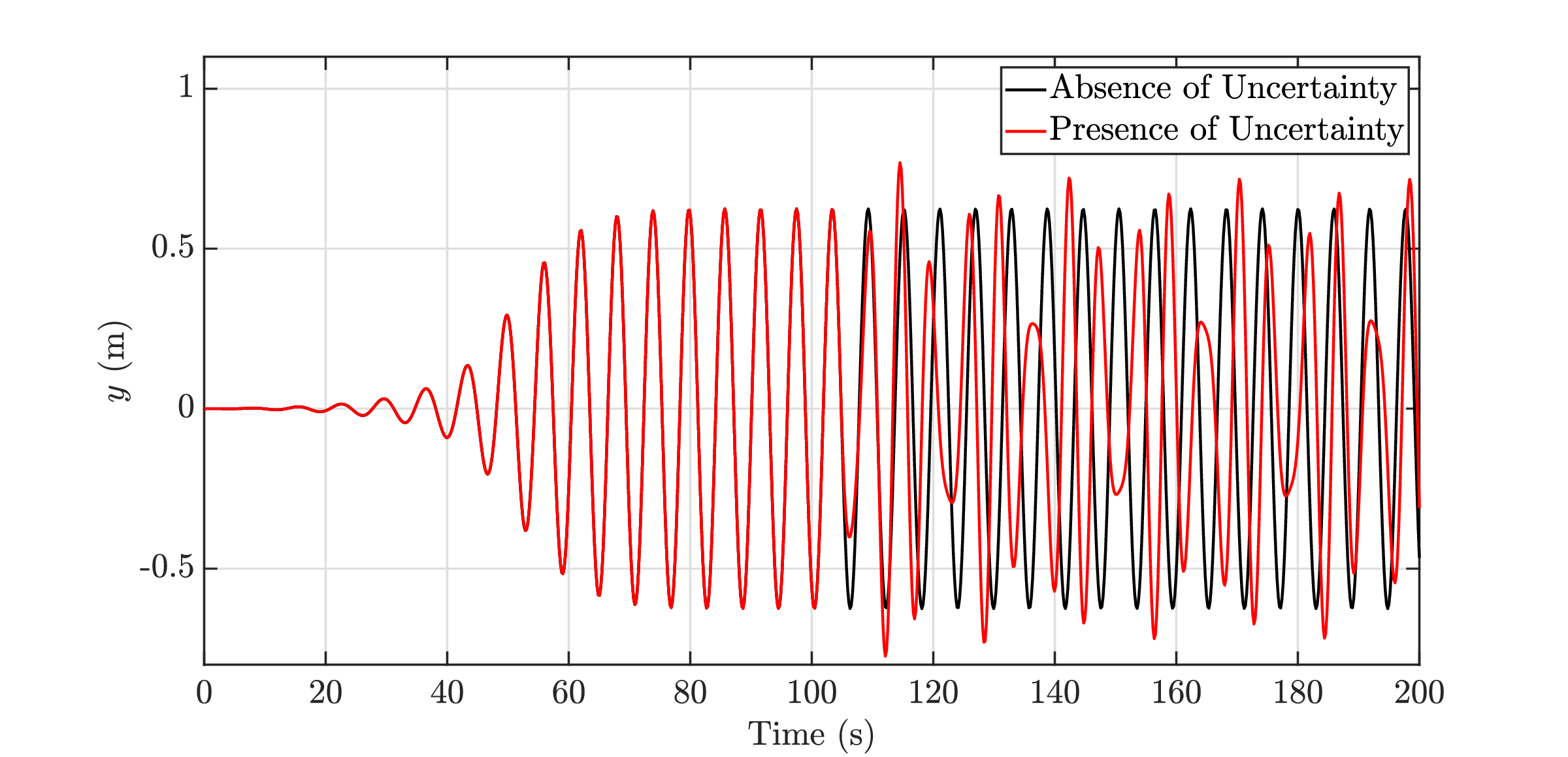}
        \label{figa_freevibunc}
        \caption{$y$ position displacement}
    \end{subfigure}%
    \begin{subfigure}[b]{0.49\textwidth}
        \centering
        \includegraphics[width=\textwidth]{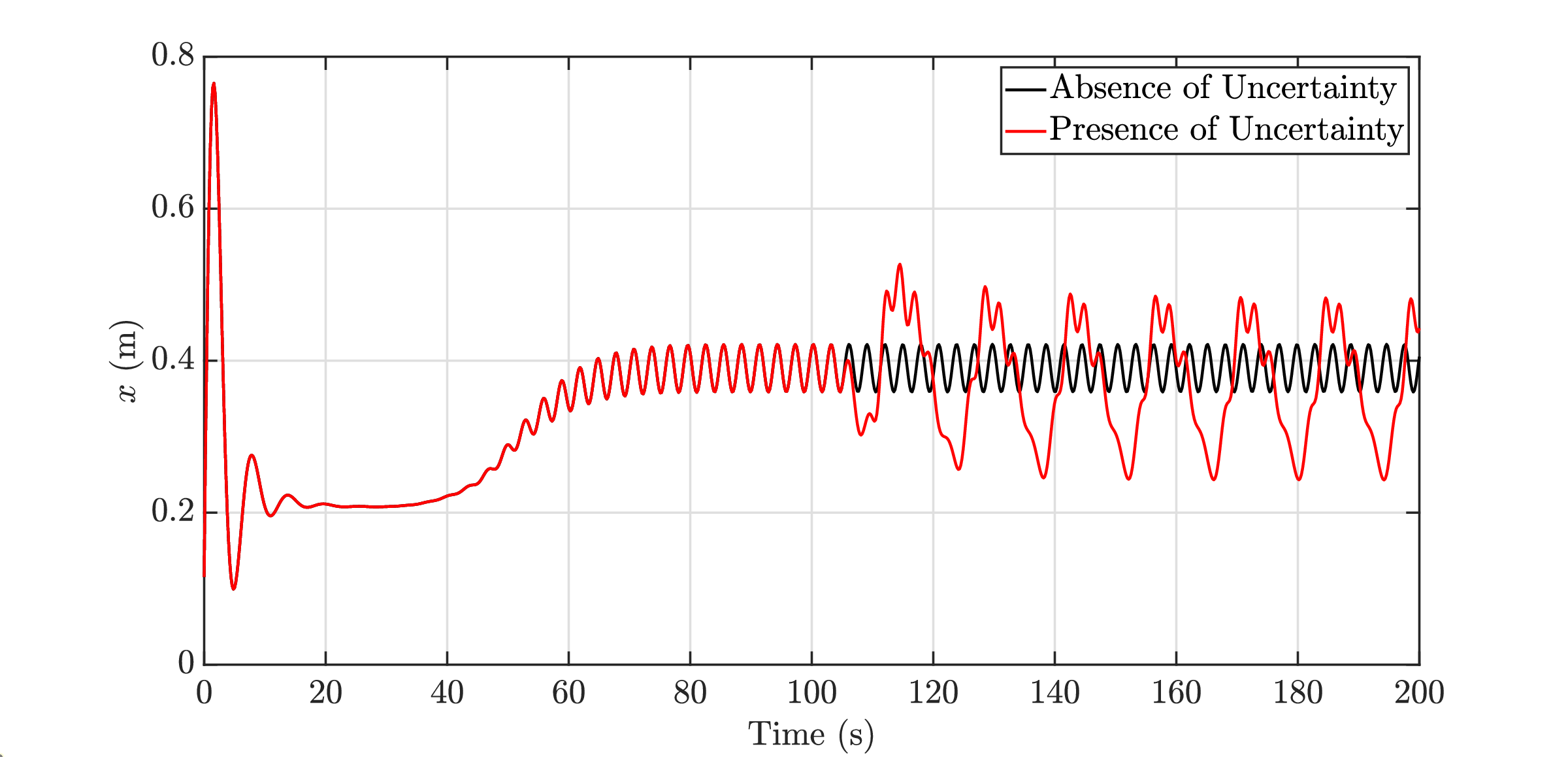}
        \label{figb_freevibunc}
        \caption{$x$ position displacement}
    \end{subfigure}%
    \caption{Free Vibration in $y$ and $x$ directions in the presence of uncertainty}
    \label{free vibration unc}
\end{figure}

The impact of applying the uncertainty term on the system's behavior can also be observed in the free vibration response. Fig.~\ref{free vibration unc} depicts the response when uncertainty term is present. After the $105^{th}$ second, the vibration amplitude varies more rapidly for both $x$ and $y$, making the suppression of VIVs and system control more challenging.

\begin{figure}[H]
  \centering
  \includegraphics[width=0.8\linewidth]{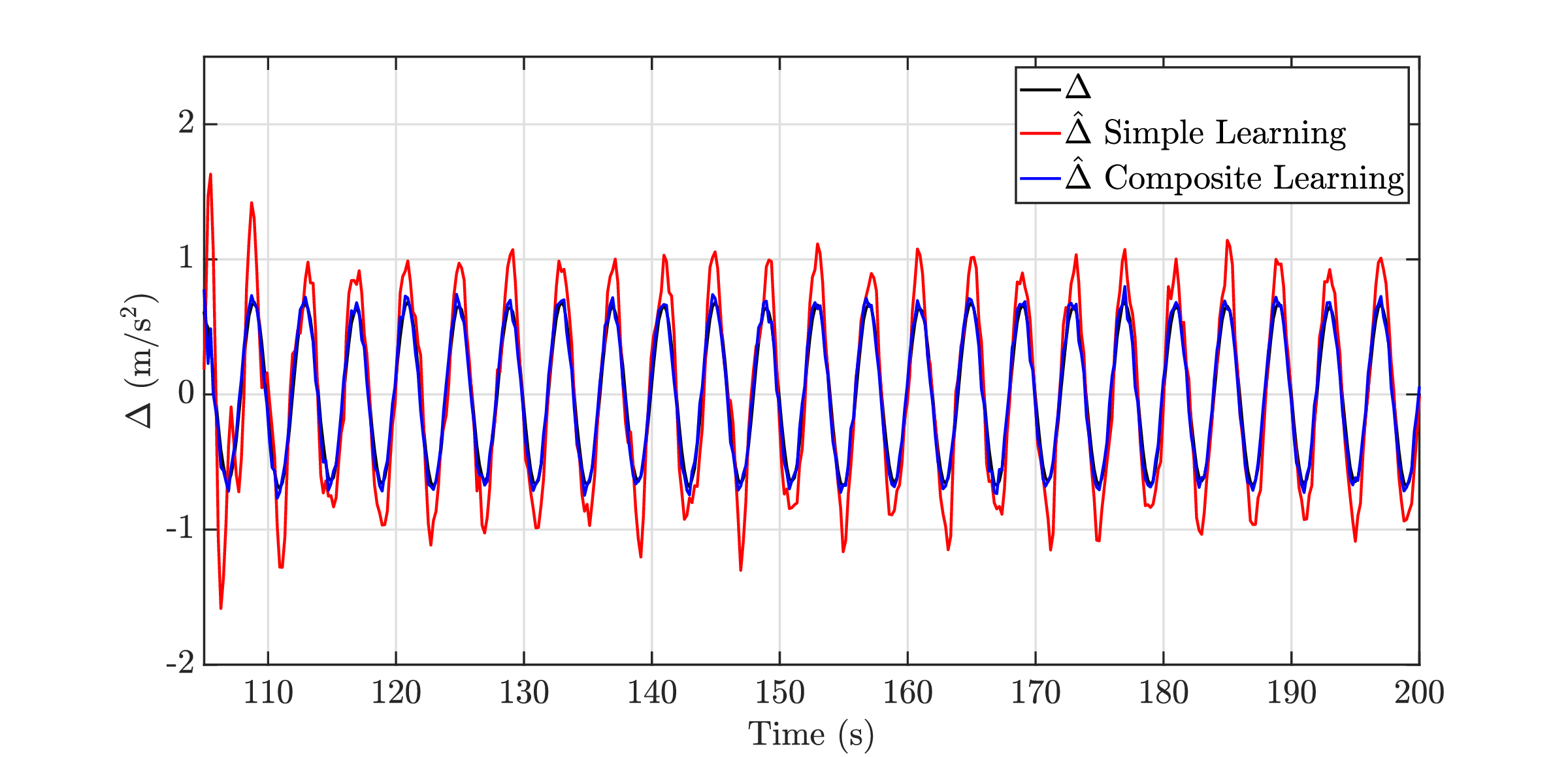}
  \caption{Uncertainty estimation by the Neural Network}
  \label{Deltahat}
\end{figure}

To assess the Neural Network's estimation of the model's uncertainty parameter, Fig.~\ref{Deltahat} is presented. It is evident that $\hat{\Delta}$ closely tracks $\Delta$ when the Composite Learning method is applied, with only minor, negligible errors at the peaks. The Simple Learning method, while less accurate than Composite Learning, still follows the general trend, though with a steady-state error. This performance distinction justifies the selection of Composite Learning for this study.

\begin{figure}[H]
    \centering
    \begin{subfigure}[b]{0.49\textwidth}
        \centering
        \includegraphics[width=\textwidth]{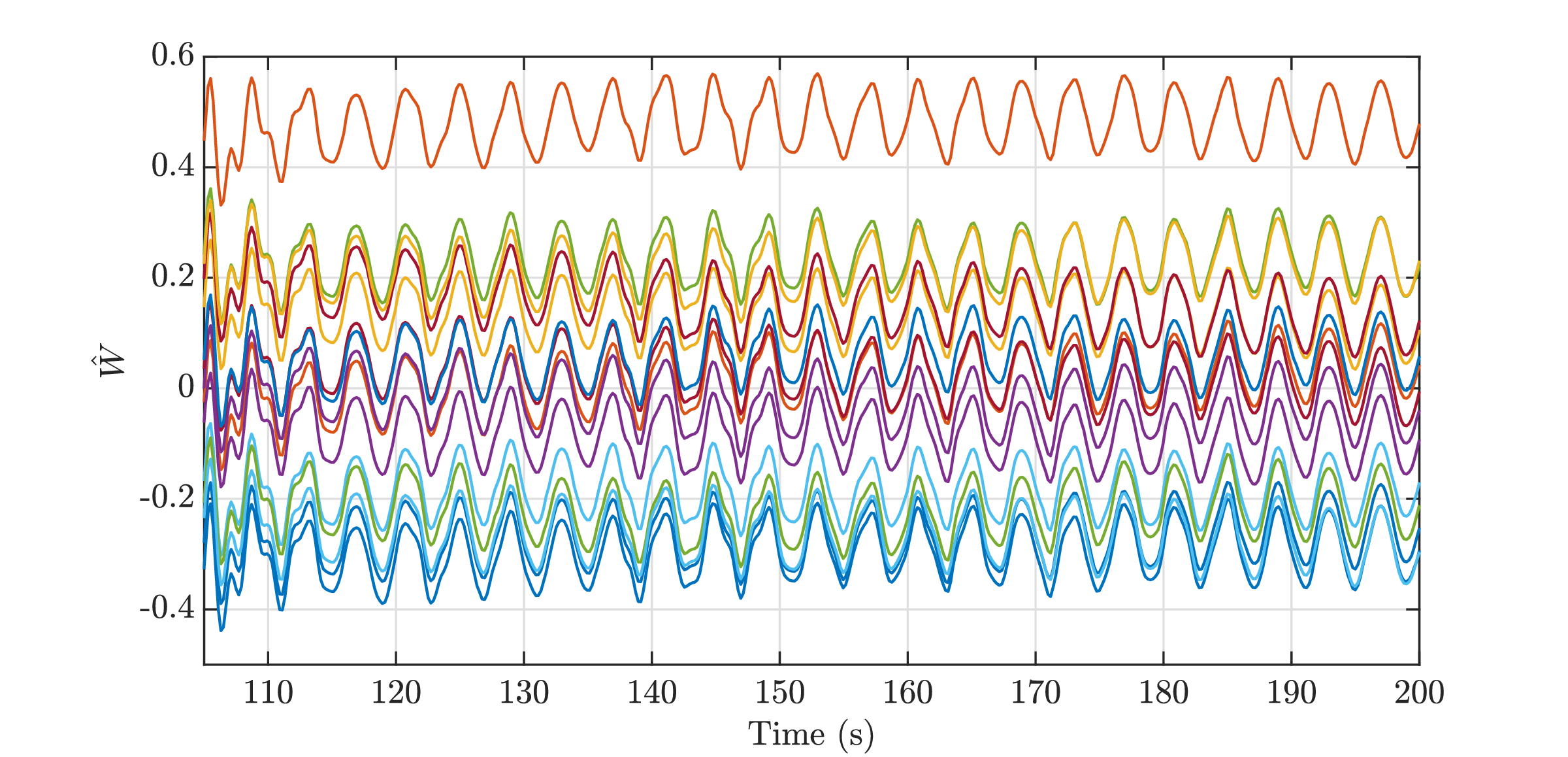}
        \caption{Simple Learning}
        \label{figa_what}
    \end{subfigure}%
    \hfill
    \begin{subfigure}[b]{0.49\textwidth}
        \centering
        \includegraphics[width=\textwidth]{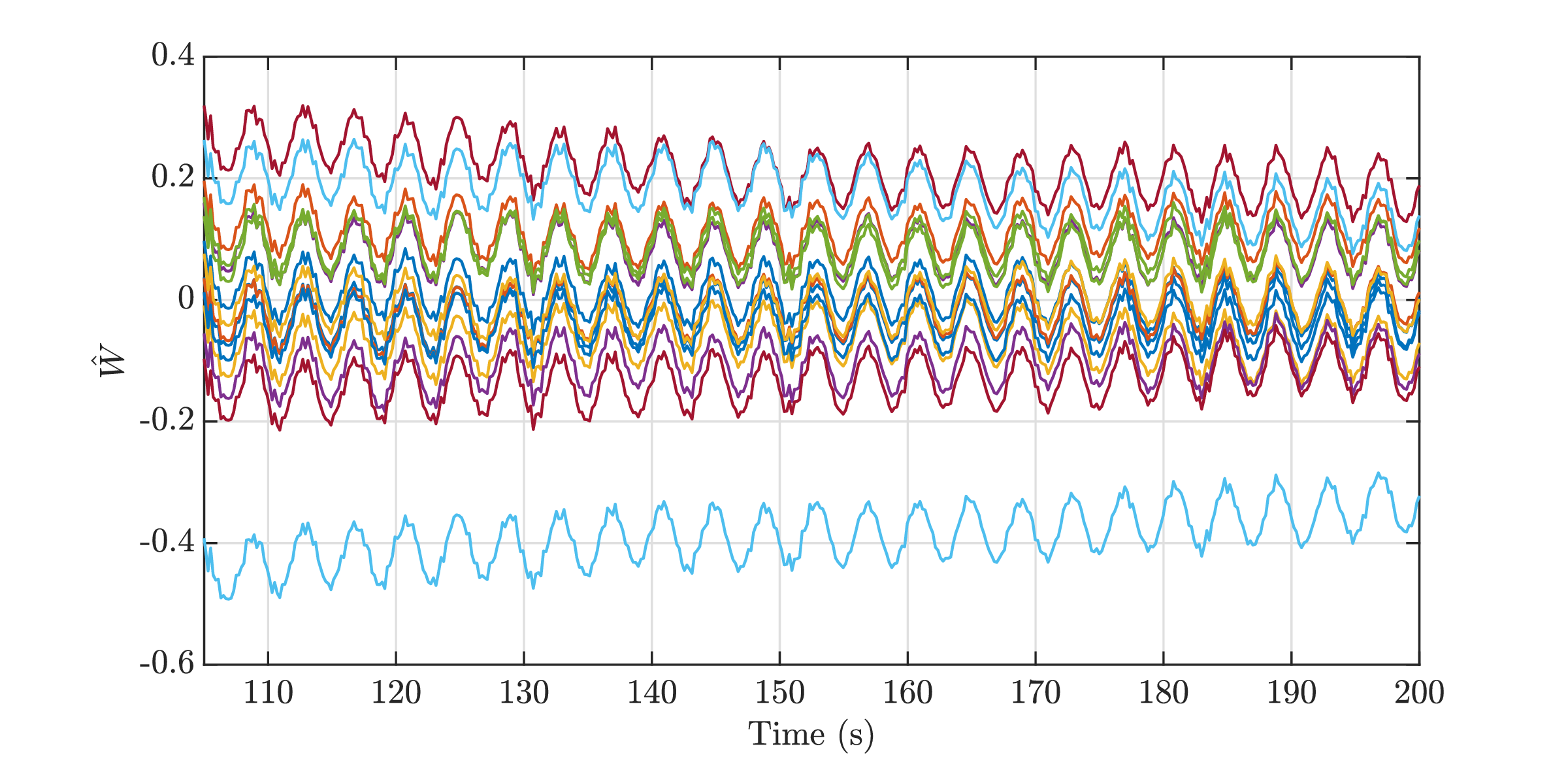}
        \caption{Composite Learning}
        \label{figb_what}
    \end{subfigure}%
    \caption{Neural Network output weights}
    \label{What unc}
\end{figure}

Finally, to evaluate the Neural Network's performance in accordance with universal approximation theorem, Fig.~\ref{What unc} is presented. According to Sec.~\ref{subsec: universal_approximation}, if the Neural Network training process is executed correctly, $\hat{W}$ should converge. Fig.~\ref{What unc} confirms that $\hat{W}$ achieves convergence in this study for both methods. However, due to the constant, time-variant sine disturbance discussed in Table~\ref{tab_uncertainties}, the convergence is to a bounded region with slight fluctuations, rather than a single point. Therefore, the Neural Network effectively identifies the unknown components of the model, contributing to enhanced control and improved suppression of VIVs.

\subsection{Evaluation of the Controller}
\label{subsec: eval controller}

The performance of the controller in suppressing VIVs is evaluated in this section. As discussed in Sec.~\ref{sec: controller formulation}, two approaches are examined: Simple Learning and Composite Learning. Both methods utilize the same values for shared parameters, as detailed in Table~\ref{tab_controlparameters}.

\begin{table}[H]
\centering
\caption{Controller Parameters}
\label{tab_controlparameters}
\small
\begin{tabularx}{\textwidth}{l X X}
    \toprule
    Description & Parameter & Value \\
    \midrule
    Learning rate of the neural network & $\Gamma$ & 1.5 \\
    Sliding surface coefficient & $\lambda$ & 12 \\
    Controller gain parameter & $k_\text{C}$ & 1.5 \\
    Sliding surface coefficient for state estimation in Composite Learning & $\lambda_\text{D}$ & 0.01 \\
    State estimation parameter in Composite Learning & $k_\text{SE}$ & 0.01 \\
    Parameter for the updating rule in Composite Learning & $k_\text{D}$ & 6 \\
    \bottomrule
\end{tabularx}
\end{table}

\begin{figure}[H]
    \centering
    \begin{subfigure}[b]{0.33\textwidth}
        \centering
        \includegraphics[width=\textwidth]{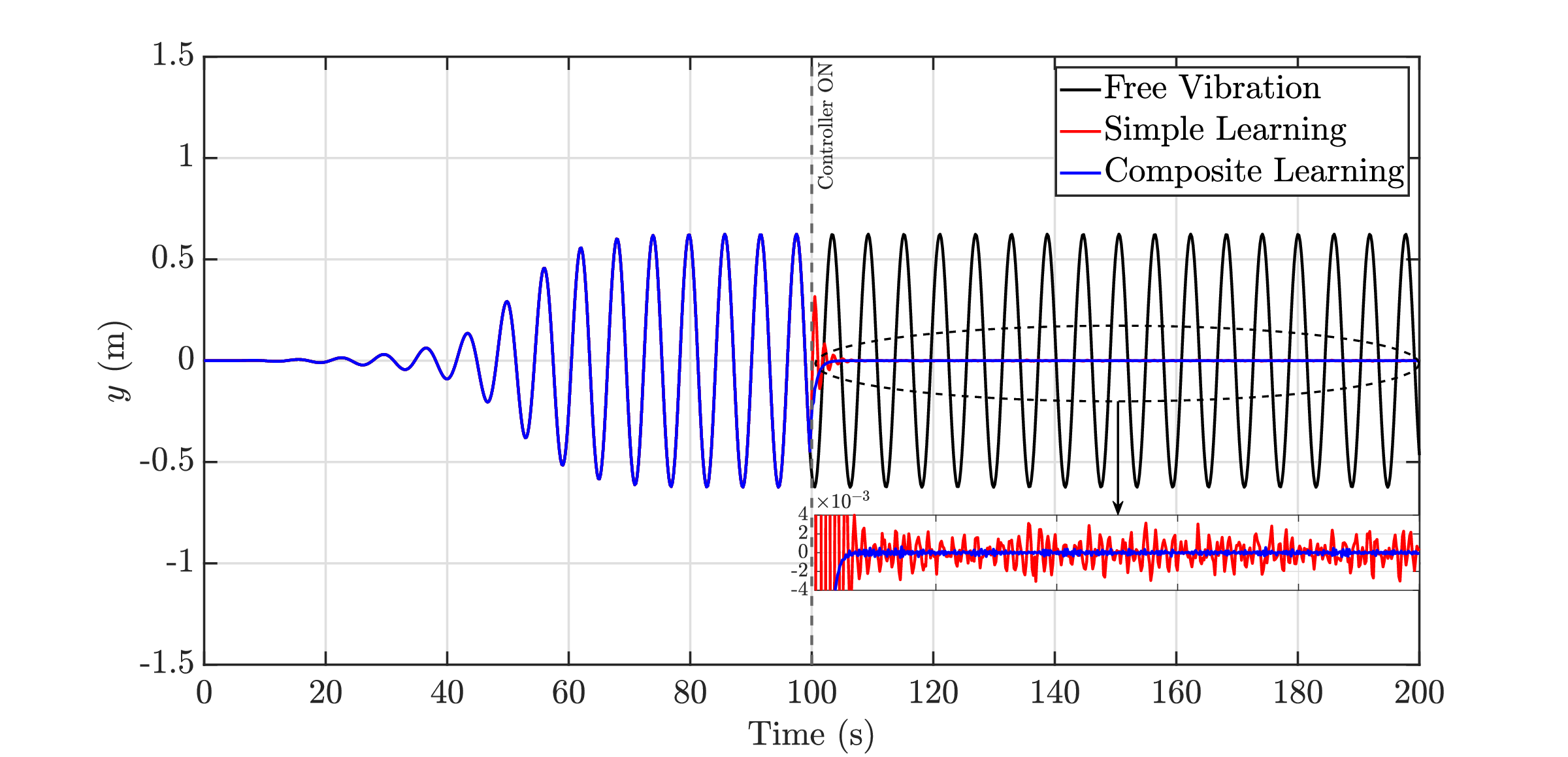}
        \caption{$y$ position displacement}
        \label{figa_y}
    \end{subfigure}%
    \hfill
    \begin{subfigure}[b]{0.33\textwidth}
        \centering
        \includegraphics[width=\textwidth]{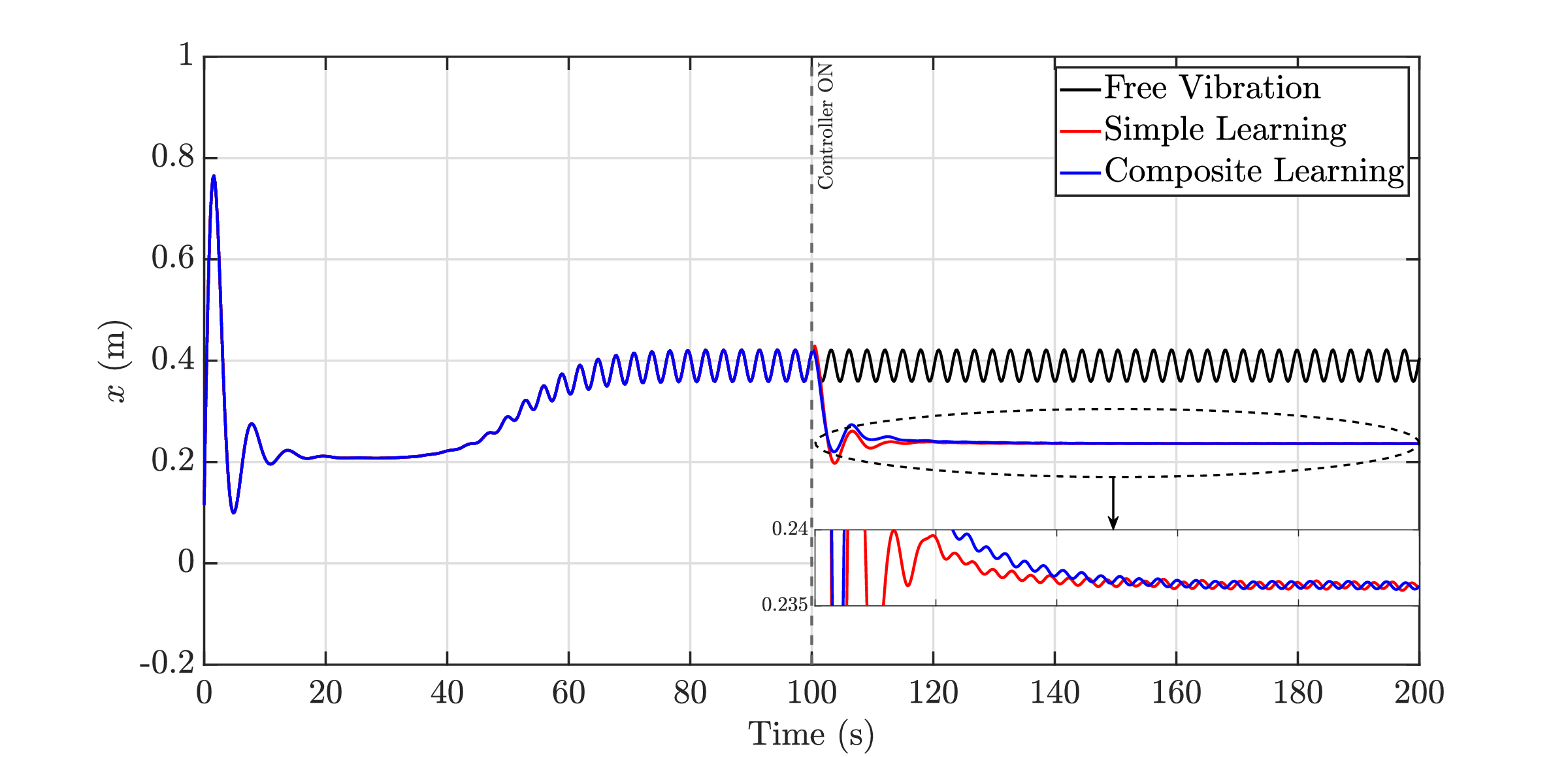}
        \caption{$x$ position displacement}
        \label{figa_x}
    \end{subfigure}%
    \hfill
    \begin{subfigure}[b]{0.33\textwidth}
        \centering
        \includegraphics[width=\textwidth]{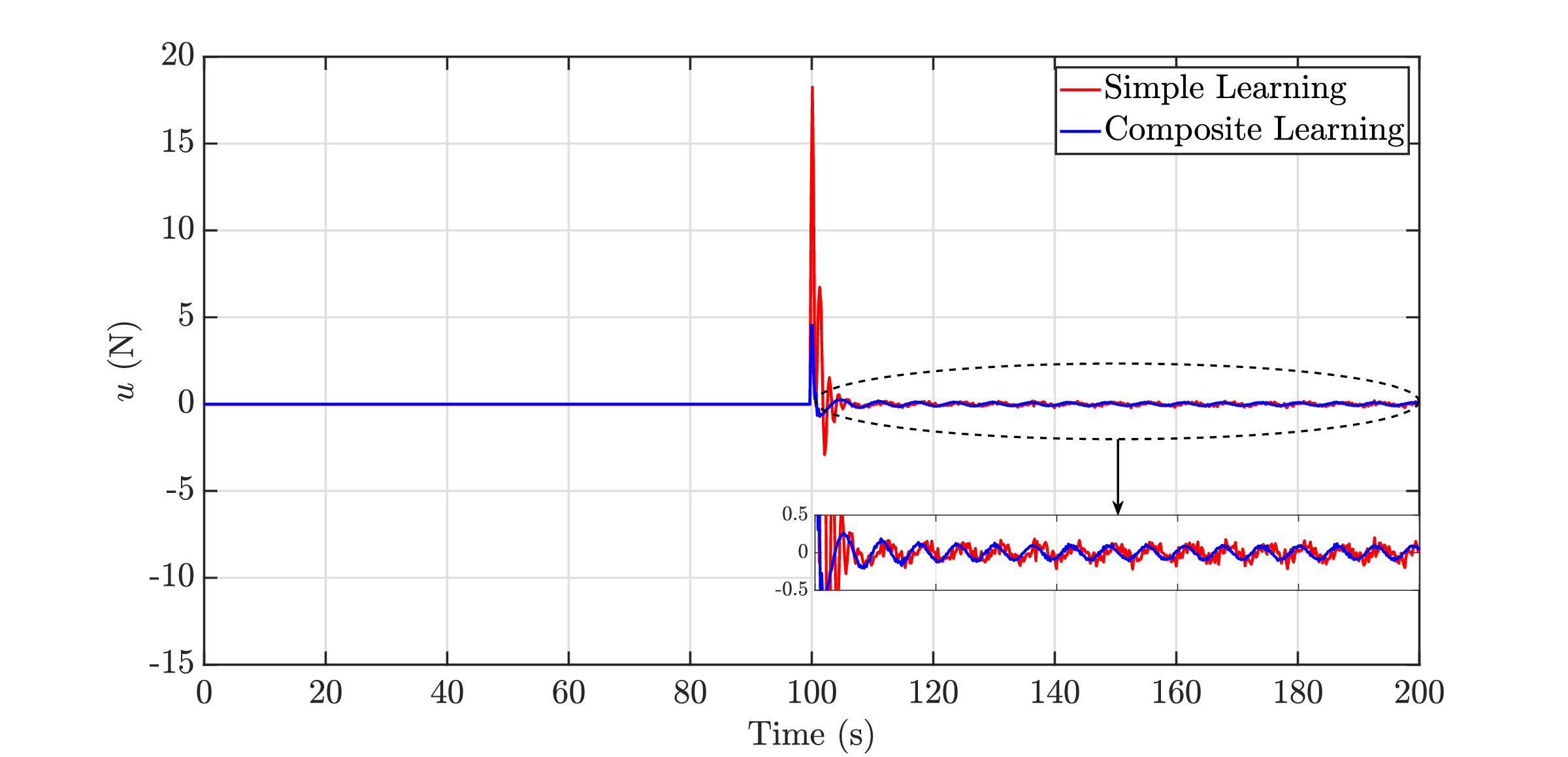}
        \caption{Control Command}
        \label{figa_u}
    \end{subfigure}
    \caption{Cylinder responses in the absence of uncertainty}
    \label{position_without_delta}
\end{figure}

Figure~\ref{position_without_delta} illustrates the cylinder's responses in the absence of the uncertainty term. As shown in Fig.~\ref{figa_y}, both methods successfully suppress vibrations in the $y$ direction. The primary difference lies in the convergence speed. With Composite Learning, the controller activates at the 100-second mark and achieves smooth convergence to zero within 2 seconds. In contrast, Simple Learning gradually reduces the vibration amplitude with some fluctuations, converging to zero after approximately 10 seconds. A closer inspection reveals that Composite Learning achieves superior performance, reducing vibration amplitude more effectively and avoiding abrupt changes, thus demonstrating smoother control. The suppressed vibration amplitude is on the order of \(10^{-3}\) for Simple Learning and \(10^{-4}\) for Composite Learning.

In the $x$ direction, as shown in Fig.~\ref{figa_x}, the controller effectively dampens vibrations despite primarily targeting the $y$ direction. The performance difference between the methods is less pronounced in this case, but Composite Learning still achieves faster convergence. A closer view shows that the final vibration amplitudes in the $x$ direction are on the order of \(10^{-4}\) for both methods.

Figure~\ref{figa_u} shows the control command. Initially, the control command for Simple Learning is approximately four times larger than that of Composite Learning. Furthermore, Simple Learning exhibits rapid fluctuations in the control command, while Composite Learning provides a smoother and more stable response.

\begin{figure}[H]
    \centering
    \begin{subfigure}[b]{0.33\textwidth}
        \centering
        \includegraphics[width=\textwidth]{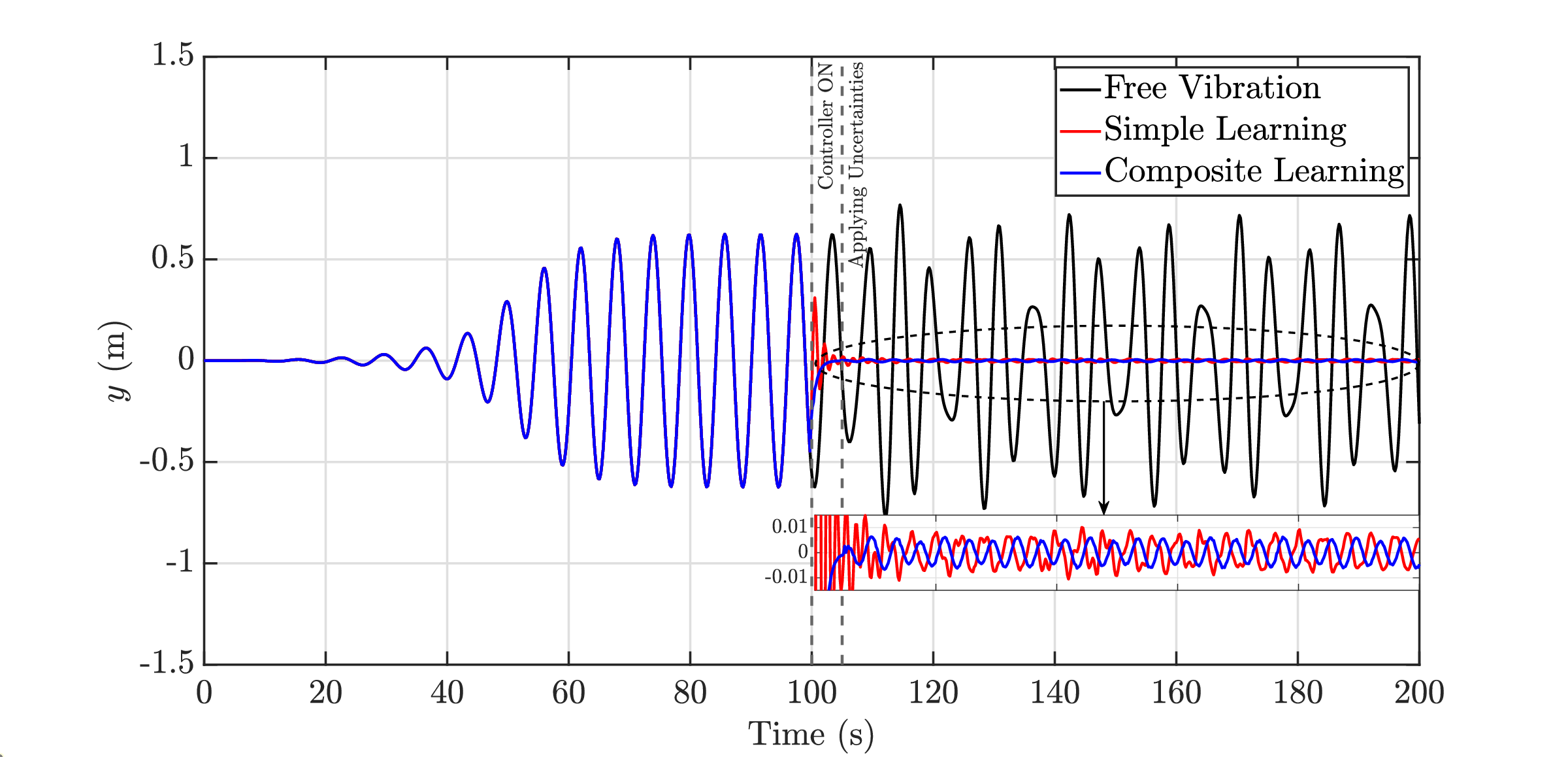}
        \caption{$y$ position displacement}
        \label{figa_yUNC}
    \end{subfigure}%
    \hfill
    \begin{subfigure}[b]{0.33\textwidth}
        \centering
        \includegraphics[width=\textwidth]{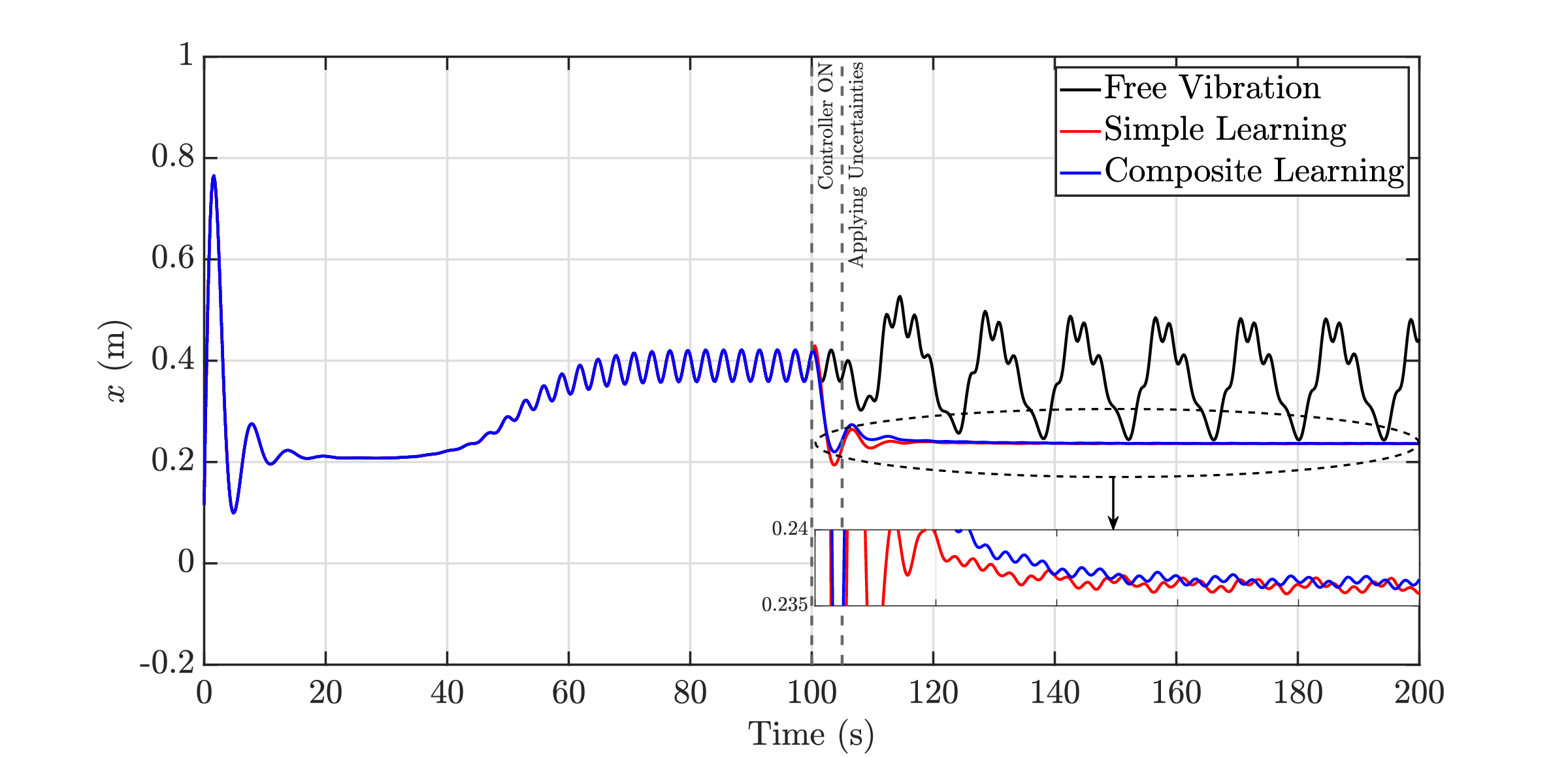}
        \caption{$x$ position displacement}
        \label{figa_xUNC}
    \end{subfigure}%
    \hfill
    \begin{subfigure}[b]{0.33\textwidth}
        \centering
        \includegraphics[width=\textwidth]{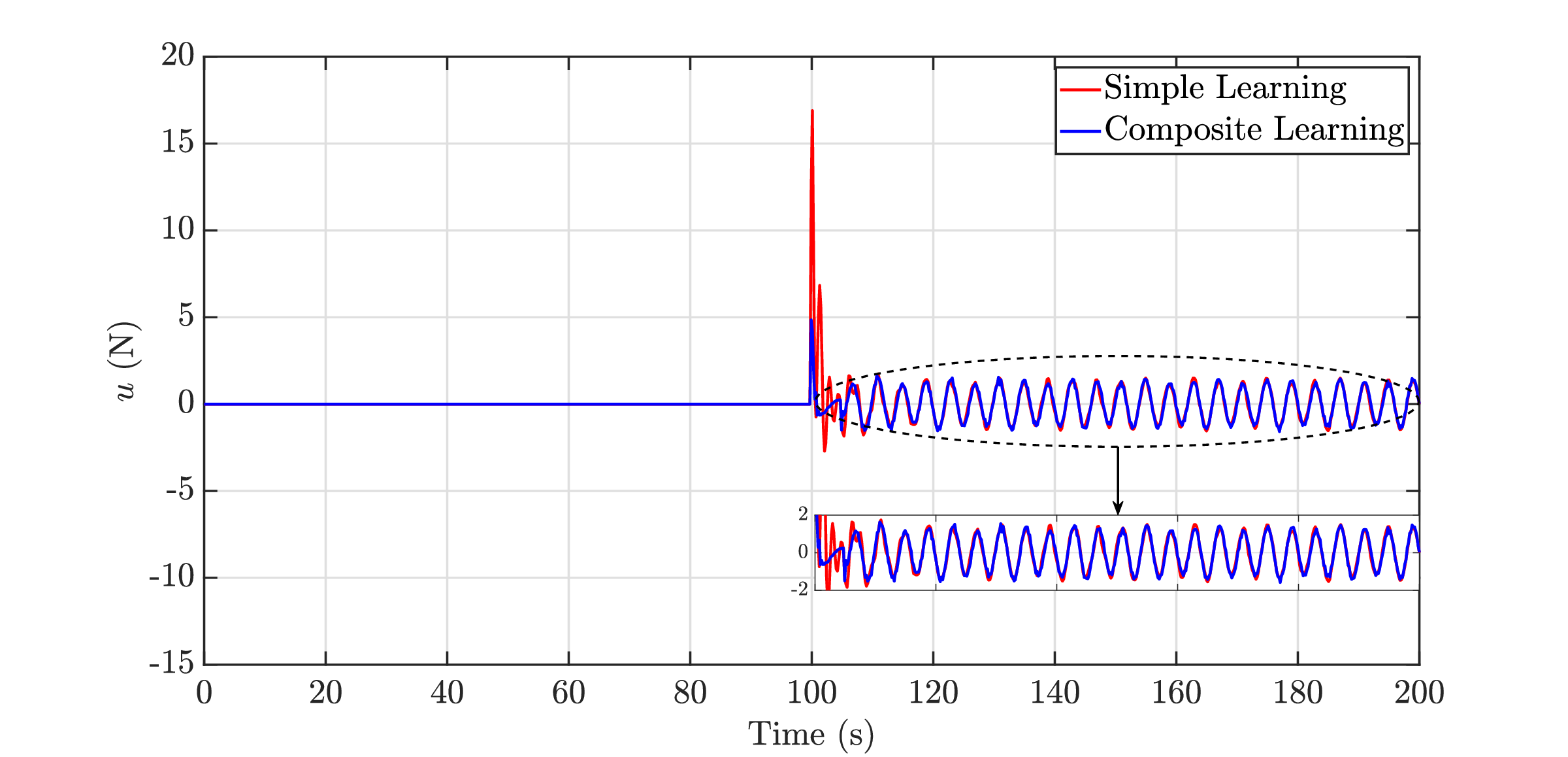}
        \caption{Control Command}
        \label{figa_uUNC}
    \end{subfigure}
    \caption{Cylinder responses in the presence of uncertainty}
    \label{position_with_delta}
\end{figure}

The cylinder's responses in the presence of uncertainty are shown in Fig.~\ref{position_with_delta}. Both methods maintain effective suppression of $y$ and $x$ vibrations. As depicted in Figs.~\ref{figa_yUNC} and \ref{figa_xUNC}, Composite Learning suppresses vibrations more promptly than Simple Learning. The vibration amplitude in the $y$ direction reduces to an order of \(10^{-3}\) for Simple Learning and \(10^{-4}\) for Composite Learning. In the $x$ direction, the vibration amplitude remains at an order of \(10^{-4}\) for both methods. These results highlight the robustness of both methods, even in the presence of uncertainty.

Figure~\ref{figa_uUNC} illustrates the control commands under uncertainty. Although the initial command for Simple Learning is slightly reduced compared to the uncertainty-free case, it remains approximately 3.5 times larger than that of Composite Learning. After a short period, both methods converge to similar control commands, as shown in the detailed view.

In summary, both Simple Learning and Composite Learning effectively suppress VIVs, as shown in Table~\ref{tab_result_compare}. Composite Learning consistently outperforms Simple Learning by achieving faster convergence and a smoother control response, even in the presence of uncertainty. These results underscore the advantages of Composite Learning in enhancing control stability and minimizing abrupt fluctuations, making it a robust approach for managing VIVs under various conditions.

\begin{table}[H]
\centering
\setlength{\extrarowheight}{0pt}
\caption{Comparison of Suppression Results}
\label{tab_result_compare}
\small
\begin{tabularx}{\textwidth}{X c c c c c}
    \toprule
    & \multicolumn{2}{c}{Absence of Uncertainty} & & \multicolumn{2}{c}{Presence of Uncertainty} \\
    \cmidrule(lr){2-3} \cmidrule(lr){5-6}
    & $y_\text{max}$ (m) & \% Suppression & & $y_\text{max}$ (m) & \% Suppression \\
    \midrule
    Free Vibration     & $6.25 \times 10^{-1}$   & -   & & $7.69 \times 10^{-1}$      & -      \\
    Simple Learning    & $3.10 \times 10^{-3}$ & 99.5  & & $1.10 \times 10^{-2}$      & 98.6      \\
    Composite Learning & $6.30 \times 10^{-4}$ & 99.9 & & $6.30 \times 10^{-3}$      & 99.2      \\
    \bottomrule
\end{tabularx}
\end{table}

\begin{figure}[H]
  \centering
  \includegraphics[width=0.8\linewidth]{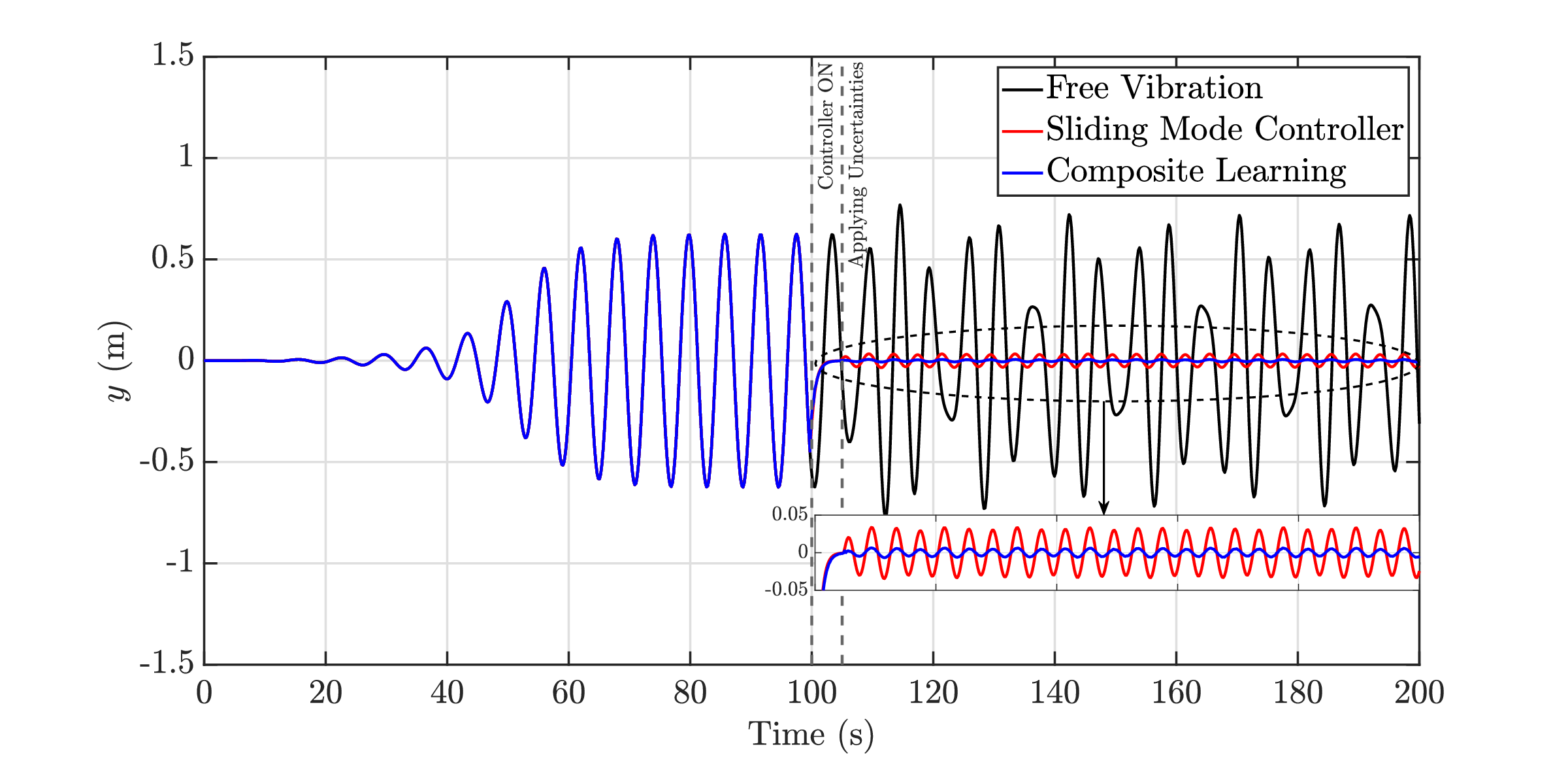}
    \caption{Comparison of cylinder vertical response for NN-VIV Suppression and SMC}
    \label{SMC compare}
\end{figure}

Due to the superior performance of Composite Learning, Fig.~\ref{SMC compare} presents a comparison between the Neural Network-based Controller and the conventional nonlinear controller, Sliding Mode Controller (SMC). By excluding the $\hat{\Delta}$ term from Eq.~\ref{control command}, the resulting control command becomes a nonlinear control command using the sliding surface, which is categorized as SMC. As illustrated in Fig.~\ref{SMC compare}, while response time and settling time for both methods are almost the same, the integration of the Neural Network to estimate the uncertainty term demonstrates its superior performance compared to the conventional SMC, with nearly 80\% more suppression of VIVs.

Despite the promising results, the cost of conducting simulations plays a crucial role in assessing the overall viability of the proposed NN-VIV Suppression method. Therefore, the simulation costs for $y$ position displacement are provided in Table~\ref{tab_cost_calc}. The state-of-the-art methods used for comparison include the Sum of Squared Errors (SSE), Mean Squared Error (MSE), Root Mean Squared Error (RMSE), and Mean Absolute Error (MAE), all of which compare the displacement results of both methods with the desired zero point. As shown in the table, Composite Learning and the integration of the Neural Network for suppressing VIVs outperform the conventional SMC method, despite the presence of significant uncertainties. The differences across all metrics indicate that NN-VIV suppression outperforms SMC by impressive percentages.

\begin{table}[H]
\centering
\setlength{\extrarowheight}{0pt}
\caption{Comparison of Simulation Cost}
\label{tab_cost_calc}
\small
\begin{tabularx}{\textwidth}{l X X X X}
    \toprule
     & SSE & MSE & RMSE & MAE \\
    \midrule
    NN-VIV Suppression & 0.2406 & 0.0005 & 0.0219 & 0.0060 \\
    Sliding Mode Controller & 0.5345 & 0.0011 & 0.0327 & 0.0221 \\
    Difference (\%) & 54.9 & 54.5 & 33.0 & 72.9 \\
    \bottomrule
\end{tabularx}
\end{table}

\section{Conclusion}
\label{sec: conclusion}

This study presents a comprehensive framework for controlling vortex-induced vibrations (VIV) in cylindrical structures. The fluid dynamics were modeled and simulated using ANSYS/Fluent, while the structural dynamics were captured in MATLAB/Simulink. A model validation study confirmed the reliability of the fluid-structure interaction model, ensuring its suitability for control design. A neural network was employed to estimate the unknown uncertainties in the system, utilizing a radial basis function (RBF) activation function with 15 neurons. The effectiveness of the neural network was validated through the convergence of output weights (\(\hat{W}\)) and the alignment of estimated uncertainty (\(\hat{\Delta}\)) with the actual uncertainty (\(\Delta\)). This enabled accurate real-time uncertainty compensation, a critical factor for robust control. Furthermore, a controllability analysis was performed to evaluate the system's ability to respond to control inputs, confirming the feasibility of achieving effective VIV suppression under various operating conditions. Two control design approaches were investigated: simple learning and composite learning. Composite learning, leveraging a state estimation function, demonstrated superior performance in terms of faster convergence to the desired state, smoother vibration suppression, and more reliable control commands compared to simple learning. This method provided a stable and efficient response, effectively addressing both structural vibrations and flow-induced dynamics. This research highlights the potential of neural networks for estimating system uncertainties unknown to the controller, addressing a critical gap in traditional VIV control methods. The proposed approach achieved a high level of VIV suppression, with 99.9\% reduction in the absence of uncertainties and 99.2\% in their presence. Additionally, the control strategy ensured prompt convergence, smooth vibration mitigation, and reliable command generation, positioning it as a robust solution for applications requiring precise and adaptive VIV control. While the results of the NN-VIV Suppression method are promising, future work could focus on further improving its efficiency by implementing a fully-tuned neural network that adjusts the weights for all layers, allowing the model to better handle more complex uncertainties, and integrating real-time updates for both the controller and neural network parameters.

\newpage
\bibliography{cas-refs}
\bibliographystyle{unsrtnat}

\end{document}